\documentclass[useAMS,usenatbib]{mn2e}
\usepackage{natbib}
\usepackage{graphicx}\usepackage{epsfig}\usepackage{epsf}
\usepackage{amsmath}\usepackage{amssymb}\usepackage{stfloats}
\title[PTF11iqb]{PTF11iqb: Cool supergiant mass loss that bridges the
  gap between Type II\lowercase{n} and normal supernovae}

\author[Smith et al.]{Nathan Smith$^{1}$\thanks{E-mail:
    nathans@as.arizona.edu}, Jon C.\ Mauerhan$^{1,2}$, S.\ Bradley
  Cenko$^3$, Mansi M.\ Kasliwal$^4$, \newauthor Jeffrey
  M. Silverman$^{5}$, Alexei V. Filippenko$^2$, Avishay Gal-Yam$^6$,
  Kelsey I.\ Clubb$^2$, \newauthor Melissa L.\ Graham$^2$, Douglas C.\
  Leonard$^{7}$, J.\ Chuck Horst$^{7}$, G.\ Grant Williams$^{1,8}$,
  \newauthor Jennifer E.\ Andrews$^1$, Shrinivas R.\ Kulkarni$^4$,
  Peter Nugent$^{2,9}$, Mark Sullivan$^{10}$, \newauthor
  Kate Maguire$^{11}$, Dong Xu$^6$, and Sagi Ben-Ami$^6$ \\
  $^{1}$Steward Observatory, University of Arizona, 933 N. Cherry
  Ave., Tucson, AZ 85721, USA \\ $^2$Department of Astronomy,
  University of California, Berkeley, CA 94720-3411, USA \\
  $^3$Astrophysics Science Division, NASA Goddard Space Flight Center,
  Mail Code 661, Greenbelt, MD 20771 \\ $^4$Astronomy Department,
  California Institute of Technology, 1200 E. California Boulevard,
  Pasadena, CA 91125, USA \\ $^5$Department of Astronomy, University
  of Texas, Austin, TX 78712, USA \\ $^6$Department of Particle Physics
  and Astrophysics, Weizmann Institute of Science, Rehovot 76100, Israel \\
  $^{7}$Department of Astronomy, San Diego State University, San
  Diego, CA 92182-1221\\
  $^8$MMT Observatory, Tucson, AZ 85721-0065, USA \\
  $^9$Computational Research Division, Lawrence Berkeley National
  Laboratory, 1 Cyclotron Road MS 50B-4206, Berkeley, CA 94720, USA \\
  $^{10}$School of Physics and Astronomy, University of Southampton,
  Southampton SO17 1BJ, UK \\ $^{11}$European Southern Observatory for
  Astronomical Research in the Southern Hemisphere (ESO),
  Karl-Schwarzschild-Str. 2, 85748 \\ $\,$ Garching b. M\"unchen, Germany}

\begin{document}

\pagerange{\pageref{firstpage}--\pageref{lastpage}} \pubyear{2012}
\maketitle
\label{firstpage}

\begin{abstract}

  The supernova (SN) PTF11iqb was initially classified as a Type~IIn
  event caught very early after explosion.  It showed narrow
  Wolf-Rayet (WR) spectral features on day 2 (as in SN~1998S and
  SN~2013cu), but the narrow emission weakened quickly and the
  spectrum morphed to resemble those of Types II-L and II-P.  At late
  times, H$\alpha$ emission exhibited a complex, multipeaked profile
  reminiscent of SN~1998S.  In terms of spectroscopic evolution, we
  find that PTF11iqb was a near twin of SN~1998S, although with a
  factor of $\sim2$--4 weaker interaction with circumstellar material
  (CSM) at early times, and stronger CSM interaction at late times.
  We interpret the spectral changes as caused by early interaction
  with asymmetric CSM that is quickly (by day 20) enveloped by the
  expanding SN ejecta photosphere, but then revealed again after the
  end of the plateau when the photosphere recedes.  The light curve
  can be matched with a simple model for weak CSM interaction (with a
  mass-loss rate of roughly 10$^{-4}$\,M$_{\odot}$ yr$^{-1}$) added to
  the light curve of a normal SN~II-P; the relatively weak CSM
  interaction allows the plateau to be seen more clearly than in other
  SNe~IIn.  This plateau requires that the progenitor had an extended
  hydrogen envelope like a red supergiant at the moment that it
  exploded, consistent with the slow progenitor wind speed ($<80$ km
  s$^{-1}$) indicated by narrow H$\alpha$ emission.  The cool
  supergiant progenitor is significant because PTF11iqb showed WR
  features in its early spectrum --- meaning that the presence of such
  WR features in an early SN spectrum does not necessarily indicate a
  WR-like progenitor.  While the late-time H$\alpha$ profile was
  multipeaked and asymmetric like that of SN~1998S, PTF11iqb's
  H$\alpha$ developed a stronger {\it redshifted} peak, so in this
  case the asymmetry cannot be blamed on dust obscuration. Instead,
  azimuthal asymmetry due to binary interaction is likely.  Overall,
  PTF11iqb bridges SNe~IIn with weaker pre-SN mass loss seen in SNe
  II-L and II-P, implying a continuum between these types.  It hints
  at episodic pre-SN mass loss on a wider scale than previously
  recognised.

\end{abstract}


\section{INTRODUCTION}

Recent years have seen the growing recognition that episodic mass loss
may be a critical ingredient in the evolution of massive stars.
Perhaps the most vivid demonstration is the class of Type~IIn
supernovae (SNe), which have narrow emission lines
\citep{schlegel90,filippenko97} arising from extremely dense
circumstellar material (CSM) that is close to the star and must have
been ejected in only the few years or decades preceding the final
explosion; see \citet{smith14} for a general review of massive-star
mass loss and its connection to SN diversity.


The synchronisation of this episodic mass loss occurring such a short
time before core collapse suggests that it is connected to some
instability in the latest stages of nuclear burning --- most likely
during Ne, O, and Si burning, but perhaps also C burning on longer
timescales \citep{qs12,sq14,sa14}.  Some SNe have associted precursor
eruptions detected in the few years before explosion
\citep{foley07,pastorello07,smith+10,mauerhan13a,corsi14,fraser13,ofek13a,ofek14a}.
Since stars may alter their structure shortly before core collapse,
this makes it challenging to connect SNe to the properties of
traditional populations of evolved stars.

Eruptions driven by the pulsational pair-instability (PPI;
\citealt{rakavy67}) have been suggested as a possible physical cause
of some of this violent late-phase mass loss (e.g.,
\citealt{woosley07}).  While the PPI could potentially be important in
the most extreme and rare events like superluminous SNe~IIn that
require ejection of 10--20\,M$_{\odot}$ (e.g.,
\citealt{smith08,smith10,ofek14b,galyam12,zhang12}), the requirements
on the very high initial mass of a PPI progenitor
($\sim100$\,M$_{\odot}$; \citealt{heger03}) as well as a preference
for low metallicity indicate that such events must be far too rare to
account for the general population of SNe~IIn
\citep{smith11,sa14,smith+14}.  Multidimensional numerical simulations
of the latest phases of nuclear burning in massive stars reveal
instabilities that may be more widespread (i.e., in lower-mass stars
of 20--30\,M$_{\odot}$), and might lead to violent eruptive mass loss
\citep{am11}.  \citet{qs12} suggested that these latest phases of Ne
and O burning might induce wave-driven mass loss that could contribute
(see also \citealt{sq14}). A number of other possibilities may exist
as well: \citet{sa14} pointed out that even if waves propagating out
from the core are insufficient to cause hydrodynamic mass loss on
their own, their dissipation in the envelope may cause a pulsation or
inflation of the star, perhaps triggering a collision with a companion
star in a binary system that had previously been safely separated.
Moreover, violent convection and mixing in the final phases may be
able to trigger explosive nuclear burning that could cause sudden
energy deposition.  In any case, the empirical fact that some SNe
experience violent eruptive mass loss in their latest burning phases
indicates that the stellar structure may be significantly modified
compared to the endpoints of one-dimensional stellar-evolution models;
\citet{sa14} proposed that this could be an extremely important
consideration in understanding core collapse (i.e., pre-explosion
disruption of the core structure might make some stars {\it easier} to
explode).


SNe~IIn contribute a fraction of 8--9\% of all core-collapse SNe
\citep{smith11} in a volume-limited sample of large star-forming
galaxies, and there are hints that they are as common or possibly more
common at lower metallicity (see \citealt{smith14}).  It is
interesting that $\sim10$\% of core-collapse SNe exhibit vivid
warnings of their impending core collapse in the form of violent fits
of dynamic mass loss, while the vast majority of normal SNe do not.
An open question is whether explosions classified as SNe~IIn are the
only massive stars that experience pre-SN instability, {\it or if they
  are just the most extreme expression of a more generic instability
  operating in a larger fraction of massive stars}.  Whereas the PPI
is indeed limited to a small fraction of the most massive stars,
wave-driven mass loss, inflation of the star that triggers collisions
with a companion star, and explosive burning may be more generic and
may operate over a wide range of mass.  Quantitative expectations of
the energy deposition and mass loss from theory are still uncertain,
but one can imagine that there is a substantial range in both.  If
there is a range in energy deposition from pre-SN instability, there
may also be a continuum in CSM density, from the most violent SNe~IIn
down to normal winds.  Anecdotally, it is already clear that SNe~IIn
exhibit a wide range of CSM mass with diverse radial density structure
and geometry.  They can eject as much as 10--20\,M$_{\odot}$ in the
cases of rare superluminous SNe~IIn (see above), 0.1--1\,M$_{\odot}$
in more common SNe~IIn \citep{kiewe12,ofek13b,taddia13}, and they even
have CSM that could arise from the dense winds of extreme red
supergiants \citep[RSGs;][]{smith09a,smith09b,ms12,stritz12}.

In this paper, we report evidence that the Type~IIn phenomenon may
extend to even lower mass-loss scales, in SNe that are recognised as
Type~IIn {\it only if they are caught early enough after explosion},
and then transition into more normal SN types as they age.  Thus, some
other ``normal'' SNe might also experience pre-SN instability akin to
SNe~IIn, but less extreme in scale or more limited in duration.  In
the title of this paper and throughout the text, we refer to
``normal'' SNe as those which are not seen to be strongly interacting
based on their visible-wavelength spectra (SNe~Ibc, IIb, II-P, II-L;
see \citealt{filippenko97} for a review), where the visual-wavelength
spectrum is dominated by the receding photosphere that is located
within the fast freely expanding SN ejecta, and not in the shock/CSM
interaction region.  Thus, normal SNe in this context exclude Types
IIn and Ibn.  Signs of weaker CSM interaction have often been seen in
X-ray and radio emission \citep{cf94,fransson96,murase14}.

\begin{figure}
\includegraphics[width=2.9in]{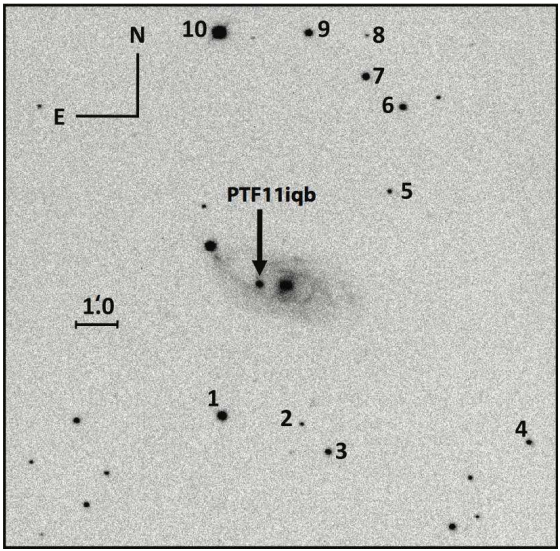}
\caption{$B$-band image of NGC 151 taken on 2011 August 5 with the Mount
  Laguna Observatory 40-inch reflecting telescope. PTF11iqb is
  indicated by the arrow. The local standards listed in
  Table~\ref{tab:stand} are marked.}
\label{fig:finder}
\end{figure}

We present visual-wavelength photometry and spectroscopy of PTF11iqb
taken from the time of first detection up to $\sim1100$ days
afterward.  PTF11iqb was discovered \citep{parrent11} at $R=16.8$ mag
by the Palomar Transient Factory (PTF; \citealt{law09,rau09}) on 2011
July 23 (UTC dates are used throughout this paper) in an inner spiral
arm of the nearby barred spiral galaxy NGC~151 (see
Figure~\ref{fig:finder}), and was initially classified as having a
Type~IIn spectrum (as we discuss below, however, the spectral
evolution became complicated).  Based on the redshift of $z=0.0125$
(3750 km s$^{-1}$), we adopt a distance of 50.4 Mpc ($m-M=33.51$ mag),
and a Milky Way line-of-sight reddening of $E(B-V)=0.0284$ mag
\citep{schlegel98}.  PTF11iqb resides in the bright inner region of
NGC~151, in the ring surrounding the central bar, at a projected
separation of $\sim30${\arcsec} ($\sim7$--8 kpc) from the host
galaxy's nucleus. Early reports indicated nondetections of PTF11iqb in
X-rays \citep{quimby11} and radio \citep{horesh11}.  Later, PTF11iqb
exhibited a marginal X-ray detection at $\sim24$ days after peak, and
then another non-detection 28 days after peak, which was suggested to
indicate a relatively low ratio of the X-ray luminosity to
visual-wavelength luminosity, as compared to other SNe~IIn
\citep{ofek13b}.  Our new observations are presented in \S 2, and the
light curve and spectral evolution are analyzed in \S 3.  We compare
PTF11iqb with SN~1998S in \S 4, and in \S 5 we present a simple CSM
interaction model.  In \S 6 we discuss the late-time H$\alpha$ and in
\S 7 we discuss the WR features seen in the early-time spectrum.
Section 8 presents an overview and summary, and we discuss PTF11iqb in
context with other SNe~IIn and normal (noninteracting) Type II (II-P
and II-L) SNe.

\begin{table}\begin{center}\begin{minipage}{2.9in}
      \caption{P48 photometry$^a$}
\scriptsize
\begin{tabular}{@{}lcc}\hline\hline
MJD     &R (mag)     &$\sigma$ (mag) \\ \hline
      55765.4      &16.862    &0.036  \\
      55765.4      &16.834    &0.039  \\
      55766.4      &16.135    &0.049  \\
      55766.4      &16.073    &0.050  \\
      55768.4      &15.608    &0.027  \\
      55768.4      &15.603    &0.050  \\
      55769.4      &15.462    &0.047  \\
      55769.4      &15.458    &0.050  \\
      55770.4      &15.370    &0.040  \\
      55770.4      &15.323    &0.057  \\
      55771.4      &15.291    &0.047  \\
      55771.4      &15.265    &0.057  \\
      55774.3      &15.218    &0.048  \\
      55774.4      &15.194    &0.054  \\
      55775.5      &15.147    &0.072  \\
      55779.4      &15.194    &0.050  \\
      55779.5      &15.178    &0.056  \\
      55780.4      &15.169    &0.054  \\
      55780.5      &15.180    &0.069  \\
      55782.4      &15.204    &0.060  \\
      55782.4      &15.193    &0.076  \\
      55783.3      &15.247    &0.025  \\
      55783.4      &15.258    &0.034  \\
      55784.3      &15.274    &0.028  \\
      55784.4      &15.274    &0.051  \\
      55793.3      &15.482    &0.041  \\
      55794.3      &15.486    &0.043  \\
      55811.3      &15.789    &0.047  \\
      55811.3      &15.776    &0.056  \\
      55812.2      &15.779    &0.044  \\
      55813.3      &15.822    &0.049  \\
      55813.4      &15.807    &0.055  \\
      55821.2      &15.909    &0.052  \\
      55821.3      &15.899    &0.058  \\
      55822.2      &15.921    &0.046  \\
      55822.3      &15.923    &0.062  \\
      55823.2      &15.921    &0.049  \\
      55823.3      &15.908    &0.062  \\
      55824.3      &15.939    &0.055  \\
      55824.3      &15.897    &0.070  \\
      55838.2      &16.054    &0.050  \\
      55842.3      &16.082    &0.039  \\
      55842.3      &16.085    &0.033  \\
      55850.2      &16.137    &0.051  \\
      55850.2      &16.134    &0.055  \\
      55851.2      &16.154    &0.053  \\
      55851.2      &16.160    &0.049  \\
      55852.2      &16.144    &0.067  \\
      55852.3      &16.159    &0.072  \\
      55853.2      &16.169    &0.066  \\
      55853.3      &16.186    &0.059  \\
      55854.2      &16.131    &0.064  \\
      55854.3      &16.137    &0.072  \\
\hline
\end{tabular}\label{tab:p48}\end{minipage}
\end{center}
$^a$Mould $R$ filter
\end{table}

\section{OBSERVATIONS}

\subsection{Palomar 48-inch Discovery and Photometry}

We obtained $R$-band images of the PTF11iqb field including NGC~151 on
2011 July 22.37 with the Palomar 48-inch telescope (P48) equipped
with the refurbished CFHT12k camera \citep{rahmer08}. Subtraction of a
stacked reference image of the field with
HOTPANTS\footnote{http://www.astro.washington.edu/users/becker/hotpants.htm}
revealed a new transient source at coordinates $\alpha =
00^{\mathrm{h}}34^{\mathrm{m}}04\fs84$, $\delta =
-09^\circ42''17\farcs9$ (J2000.0), with an astrometric
uncertainty (relative to the USNO-B1 catalog; \citealt{monet03}) of
$\pm$150\,mas in each coordinate.  The transient was discovered $30$
hours later by Oarical, an autonomous software framework of the PTF
collaboration \citep{bloom11}.  It was classified correctly as a
transient source (as opposed to a variable star), was further
classified as a SN or nova, and was given the name PTF11iqb.  No
source was detected at this location with P48 in an image taken on
2011 Jul 17.5, to a 3$\sigma$ limit of $R \approx
18.5$ mag (see Figures~\ref{fig:photUVOT} and \ref{fig:phot}).

Our photometric pipeline has been used in many previous PTF papers
(e.g., \citealt{firth14,ofek12,ofek14a,pan14,laher14}), and is based
on image subtraction. We construct a deep reference image from data
prior to the SN explosion, register this reference to each image
containing the SN light, match the point spread functions (PSFs),
perform the image subtraction, and then measure the SN flux using PSF
photometry on the difference images. The PSF is determined using
isolated stars in the unsubtracted images, and the image subtraction
uses a nonparametric pixelised kernel (similar to that in
\citealt{bramich08}). The average PSF is then fit at the position of
the SN event weighting each pixel according to Poisson statistics,
yielding a SN flux and flux uncertainty.  We flux calibrate our P48
light curve to the Sloan Digital Sky Survey (SDSS; \citealt{york00})
Data Release 10 (DR10; \citealt{aaa+14}). The resulting magnitudes are
given in Table~\ref{tab:p48} and are shown in
Figures~\ref{fig:photUVOT} and \ref{fig:phot}.

\subsection{Palomar 60-inch Photometry}

\begin{figure*}
\includegraphics[width=7.0in]{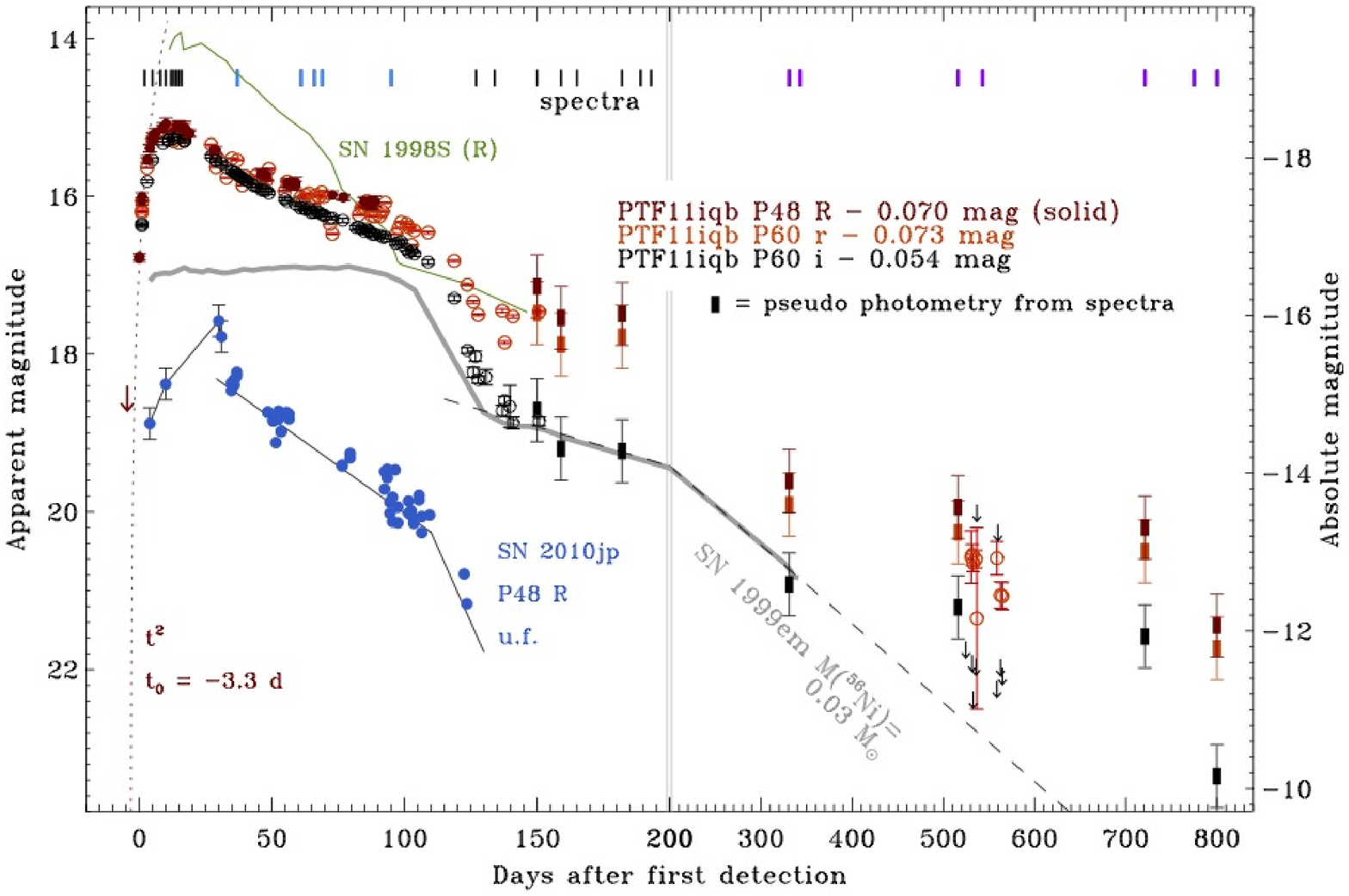}
\caption{Apparent and absolute magnitude light curve for PTF11iqb,
  with P48 $R$ in solid red dots, and with P60 $r^{\prime}$ and
  $i^{\prime}$ shown with unfilled circles in red-orange and
  black, respectively (see Tables \ref{tab:p48} and \ref{tab:p60}).
  The arrows at late times are P60 upper limits, whereas the solid
  rectangles at late times (after day $\sim150$) represent pseudo-photometry 
  estimated from spectra as described in the text.  All
  photometry for PTF11iqb in the three filters has been corrected for
  Milky Way reddening of 0.070, 0.073, and 0.054 mag, for $R$,
  $r^{\prime}$, and $i^{\prime}$, respectively.  Epochs when we
  obtained visual-wavelength spectra are noted with coloured hash-marks
  at the top (see Table 3 and Figure~\ref{fig:spec}).  For reference,
  we also show $R$-band light curves of the normal SN~II-P 1999em
  \citep{leonard02}, the unusual SN~IIn 2010jp \citep{smith12}, and
  SN~1998S \citep{fassia00}, scaled as they would appear if they were
  at the same distance as PTF11iqb. The dashed line is representative
  of the $^{56}$Co decay rate, matched to SN~1999em.  Note that the
  time axis is not uniform; we include a change in scale at 200 days
  to better facilitate the display of the early-time plateau phase on
  the same plot as the more sparsely sampled late-time decay.}
\label{fig:phot}
\end{figure*}

Upon discovery of PTF11iqb, the field was automatically inserted into
the queue of the robotic Palomar 60-inch telescope (P60;
\citealt{cenko06}) for multifilter ($r^{\prime}$ and $i^{\prime}$)
follow-up observations.  Images were processed using a custom
pipeline, and subtracted from reference images obtained in 2014 July
using HOTPANTS.  The resulting subtracted images were photometrically
calibrated using nearby point sources from the SDSS DR10 \citep{aaa+14}), 
so reported magnitudes are on the
AB system \citep{og83}.  A log of P60 photometry is provided
in Table~\ref{tab:p60}.  The light curve, combining the P48 $R$-band
and the P60 $r^{\prime}$ and $i^{\prime}$ photometry, is shown in
Figures~\ref{fig:photUVOT} and \ref{fig:phot}.

The general rate of fading in P60 $ri$ data is very consistent with
the P48 $R$-band data.  However, there are some differences that can
mostly be attributed to different amounts of H$\alpha$ emission in the
$r$ (Sloan) and $R$ (Mould) filters, and none in the $i$ filter.  We
also include a number of $i$-band upper limits at late times around
days 500--600.

\begin{table*}\begin{center}\begin{minipage}{4.3in}
      \caption{P60 $r$ and $i$ photometry of PTF11iqb$^a$}
\scriptsize
\begin{tabular}{@{}lccclccc}\hline\hline
MJD &Filt.      &mag    &$\sigma$    &MJD    &Filt. &mag     &$\sigma$ \\ \hline
55766.449	&	r	&	16.134	&	0.004	&	55768.358	&	i	&	15.948	&	0.076	\\
55770.355	&	r	&	15.347	&	0.004	&	...	        &	i	&	...	&	...	\\
55774.344	&	r	&	15.230	&	0.004	&	55774.343	&	i	&	15.445	&	0.035	\\
55776.338	&	r	&	15.236	&	0.004	&	55776.337	&	i	&	15.460	&	0.069	\\
55776.340	&	r	&	15.217	&	0.004	&	55776.339	&	i	&	15.464	&	0.053	\\
55778.331	&	r	&	15.374	&	0.004	&	55778.330	&	i	&	15.486	&	0.107	\\
55778.334	&	r	&	15.255	&	0.004	&	55778.333	&	i	&	15.441	&	0.067	\\
55780.330	&	r	&	15.396	&	0.003	&	...     	&	i	&	...  	&	...  	\\
55782.321	&	r	&	15.357	&	0.004	&	55782.320	&	i	&	15.460	&	0.063	\\
55782.323	&	r	&	15.306	&	0.004	&	55782.322	&	i	&	15.473	&	0.097	\\
55792.482	&	r	&	15.417	&	0.004	&	55792.481	&	i	&	15.720	&	0.097	\\
55794.300       &       r       &       15.557  &       0.004	&	...     	&	i	&	...  	&	...  	\\
55794.300       &       r       &       15.704  &       0.004	&	...     	&	i	&	...  	&	...  	\\
55796.282	&	r	&	15.633	&	0.004	&	55796.281	&	i	&	15.763	&	0.049	\\
55798.300       &       r       &       15.839  &       0.004	&	...     	&	i	&	...  	&	...  	\\
55800.373	&	r	&	15.593	&	0.004	&	55800.372	&	i	&	15.825	&	0.046	\\
55801.400	&	r	&       15.779  &       0.005	&	...     	&	i	&	...  	&	...  	\\
55802.300	&	r	&       15.616  &       0.004	&	...     	&	i	&	...  	&	...  	\\
55803.264	&	r	&	15.821	&	0.004	&	55803.263	&	i	&	15.948	&	0.040	\\
55804.261	&	r	&	15.936	&	0.004	&	55804.260	&	i	&	15.932	&	0.070	\\
55805.259	&	r	&	15.815	&	0.004	&	55805.258	&	i	&	15.986	&	0.084	\\
55806.256	&	r	&	15.842	&	0.005	&	...	        &	i	&	...	&	...	\\
...     	&	r	&	...	&	...	&	55807.254	&	i	&	15.935	&	0.049	\\
55808.251	&	r	&	15.888	&	0.054	&	55808.250	&	i	&	16.033	&	0.066	\\
55809.400	&	r	&       15.807  &       0.005	&	...	        &	i	&	...	&	...	\\
55811.244	&	r	&	15.904	&	0.005	&	55811.243	&	i	&	16.005	&	0.058	\\
55812.239	&	r	&	15.965	&	0.005	&	55812.238	&	i	&	16.090	&	0.076	\\
55813.200	&	r	&       15.881  &       0.006	&	...	        &	i	&	...	&	...	\\
55814.233	&	r	&	15.725	&	0.007	&	55814.232	&	i	&	16.151	&	0.042	\\
55820.257	&	r	&	16.005	&	0.006	&	55820.256	&	i	&	16.253	&	0.059	\\
55821.216	&	r	&	15.887	&	0.006	&	55821.215	&	i	&	16.248	&	0.066	\\
55827.229	&	r	&	16.107	&	0.005	&	55827.228	&	i	&	16.344	&	0.061	\\
55828.223	&	r	&	16.060	&	0.005	&	55828.222	&	i	&	16.379	&	0.052	\\
55829.200	&	r	&       16.036  &       0.006	&	...	        &	i	&	...	&	...	\\
55830.200	&	r	&       16.265  &       0.006	&	...	        &	i	&	...	&	...	\\
55831.188	&	r	&	16.278	&	0.158	&	55831.187	&	i	&	16.381	&	0.050	\\
55832.211	&	r	&	16.165	&	0.006	&	55832.210	&	i	&	16.341	&	0.063	\\
55833.200       &	r	&	16.080  &	0.006	&	55833.181	&	i	&	16.326	&	0.048	\\
55834.179	&	r	&	16.1220	&	0.006	&	55834.178	&	i	&	16.380	&	0.086	\\
55835.177	&	r	&	16.076	&	0.005	&	55835.176	&	i	&	16.358	&	0.048	\\
55837.300	&	r	&       16.425  &       0.005	&	...	        &	i	&	...	&	...	\\
55838.168	&	r	&	16.551	&	0.007	&	55838.167	&	i	&	16.616	&	0.050	\\
...     	&	r	&	...  	&	...	&	55847.144	&	i	&	16.598	&	0.058	\\
55848.141	&	r	&	16.310	&	0.008	&	55848.140	&	i	&	16.524	&	0.047	\\
55849.137	&	r	&	16.219	&	0.008	&	55849.136	&	i	&	16.594	&	0.062	\\
55850.139	&	r	&	16.210	&	0.006	&	55850.138	&	i	&	16.639	&	0.058	\\
55851.133	&	r	&	16.135	&	0.006	&	55851.132	&	i	&	16.597	&	0.085	\\
55852.129	&	r	&	16.176	&	0.006	&	55852.128	&	i	&	16.630	&	0.064	\\
55852.200	&	r	&       16.208  &       0.005	&	...	        &	i	&	...	&	...	\\
55853.130	&	r	&	16.323	&	0.006	&	55853.129	&	i	&	16.640	&	0.067	\\
55854.129	&	r	&	16.330	&	0.006	&	55854.128	&	i	&	16.687	&	0.086	\\
55856.100	&	r	&       16.324  &       0.006	&	...	        &	i	&	...	&	...	\\
55857.100	&	r	&       16.269  &       0.006	&	...	        &	i	&	...	&	...	\\
55858.100	&	r	&       16.154  &       0.006	&	...	        &	i	&	...	&	...	\\
...     	&	r	&	...	&	...	&	55859.123	&	i	&	16.813	&	0.053	\\
55862.113	&	r	&	16.555	&	0.007	&	55862.112	&	i	&	16.852	&	0.062	\\
55863.111	&	r	&	16.481	&	0.007	&	55863.110	&	i	&	16.792	&	0.041	\\
55864.100	&	r	&       16.401  &       0.007	&	...	        &	i	&	...	&	...	\\
55866.110	&	r	&	16.422	&	0.007	&	55866.109	&	i	&	16.868	&	0.069	\\
55867.100	&	r	&       16.467  &       0.009	&	...	        &	i	&	...	&	...	\\
55868.105	&	r	&	16.693	&	0.008	&	55868.104	&	i	&	16.952	&	0.060	\\
55869.100	&	r	&       16.509  &       0.012	&	...	        &	i	&	...	&	...	\\
55874.297	&	r	&	16.533	&	0.014	&	55874.296	&	i	&	17.083	&	0.065	\\
55884.100	&	r	&       16.892  &       0.012	&	...	        &	i	&	...	&	...	\\
55889.112	&	r	&	17.194	&	0.010	&	55889.111	&	i	&	18.203	&	0.069	\\
55891.296	&	r	&	17.411	&	0.016	&	...	        &	i	&	...	&	...	\\
55893.157	&	r	&	17.576	&	0.016	&	55893.156	&	i	&	18.604	&	0.075	\\
...     	&	r	&	...	&	...	&	55895.094	&	i	&	18.620	&	0.074	\\
55902.247	&	r	&	17.524	&	0.024	&	55902.245	&	i	&	18.907	&	0.081	\\
55903.112	&	r	&	17.927	&	0.017	&	55903.110	&	i	&	18.836	&	0.058	\\
...     	&	r	&	...	&	...	&	55904.109	&	i	&	18.903	&	0.078	\\
...     	&	r	&	...	&	...	&	55905.105	&	i	&	(19.76)	&	...	\\
55906.276	&	r	&	17.595	&	0.020	&	55906.274	&	i	&	19.040	&	0.095	\\
55916.133	&	r	&	17.537	&	0.016	&	55916.131	&	i	&	19.120	&	0.099	\\
...     	&	r	&	...	&	...	&	56290.090	&	i	&	(21.69)	&	...	\\
56296.100	&	r	&	20.647	&	0.332	&	...	        &	i	&	...	&	...	\\
56297.140	&	r	&	20.617	&	0.131	&	56297.139	&	i	&	(21.86)	&	...	\\
56298.115	&	r	&	20.723	&	0.109	&	56298.113	&	i	&	(22.33)	&	...	\\
56301.084	&	r	&	20.673	&	0.111	&	56301.081	&	i	&	(21.91)	&	...	\\
56302.107	&	r	&	21.423	&	1.151	&	56302.106	&	i	&	(19.96)	&	...	\\
56324.100       &	r	&	20.660	&	0.211	&	56324.128	&	i	&	(22.19)	&	...	\\
...	        &	r	&	...	&	...	&	56325.129	&	i	&	(20.21)	&	...	\\
56328.115	&	r	&	21.129	&	0.165	&	56328.113	&	i	&	(21.92)	&	...	\\
56330.112	&	r	&	21.142	&	0.173	&	56330.110	&	i	&	(22.03)	&	...	\\
\hline
\end{tabular}\label{tab:p60}\end{minipage}
\end{center}
$^a$Magnitudes in parenthesis indicate upper limits.
\end{table*}

\begin{table*}\begin{center}\begin{minipage}{3.9in}
      \caption{MLO Magnitudes of Local Standards}
\scriptsize
\begin{tabular}{@{}lcccccc}\hline\hline
Star  &{$B$ ($\sigma_B$)} &{$V$ ($\sigma_V$)} &{$R$ ($\sigma_R$)} &{$I$ ($\sigma_I$)}  \\ \hline
1 & 13.822(0.040) & 13.258(0.022) & 12.873(0.033) & 12.563(0.033)\\  
2 & 17.192(0.052) & 16.319(0.027) & 15.709(0.039) & 15.192(0.040)\\  
3 & 15.522(0.040) & 14.898(0.022) & 14.408(0.033) & 14.019(0.033)\\  
4 & 16.091(0.040) & 14.796(0.022) & 13.859(0.033) & 13.129(0.033)\\  
5 & 16.856(0.045) & 15.714(0.023) & 14.949(0.035) & 14.250(0.034)\\  
6 & 15.253(0.040) & 14.511(0.022) & 14.025(0.033) & 13.591(0.033)\\  
7 & 14.883(0.040) & 14.214(0.022) & 13.823(0.033) & 13.446(0.033)\\  
8 & 17.893(0.065) & 16.660(0.029) & 15.858(0.042) & 15.055(0.042)\\  
9 & 14.702(0.040) & 14.097(0.022) & 13.734(0.033) & 13.339(0.033)\\  
10& 11.928(0.040) & 11.416(0.022) & 11.093(0.033) & 10.778(0.033)\\  
\hline
\end{tabular}\label{tab:stand}\end{minipage}
\end{center}
\end{table*}

\begin{table*}\begin{center}\begin{minipage}{3.9in}
  \caption{MLO Photometric Observations of SN PTF11iqb}
\scriptsize
\begin{tabular}{@{}lcccccc}\hline\hline
UT Date$^{a}$ &Day$^{b}$  &MJD  &{$B$ ($\sigma_B$)} &{$V$ ($\sigma_V$)} &{$R$ ($\sigma_R$)} &{$I$ ($\sigma_I$)} \\ \hline
2011-08-03 &  11.06 & 55776.46 & ...           & ...           & 15.049(0.019) & 15.031(0.014)\\ 
2011-08-06 &  14.07 & 55779.47 & 15.334(0.064) & 15.260(0.025) & 15.051(0.082) & 14.949(0.034)\\ 
2011-09-01 &  39.99 & 55805.39 & ...           & ...           & 15.619(0.064) & ...          \\ 
2011-09-17 &  56.02 & 55821.42 & 17.036(0.045) & 16.447(0.097) & 15.902(0.036) & 15.623(0.039)\\ 
2011-09-23 &  61.95 & 55827.35 & 17.159(0.030) & 16.546(0.052) & 15.998(0.040) & 15.748(0.025)\\ 
2011-09-28 &  67.00 & 55832.41 & 17.256(0.069) & 16.518(0.042) & 15.976(0.035) & 15.727(0.033)\\ 
2011-10-19 &  87.91 & 55853.31 & 17.582(0.037) & 16.877(0.063) & 16.160(0.055) & 15.985(0.033)\\ 
2011-10-27 &  95.88 & 55861.28 & 17.738(0.075) & 16.933(0.063) & 16.258(0.037) & 16.098(0.041)\\ 
2011-11-02 & 101.86 & 55867.26 & 17.790(0.058) & 17.073(0.041) & 16.369(0.046) & 16.190(0.027)\\ 
2011-11-15 & 114.89 & 55880.29 & 18.130(0.041) & 17.525(0.068) & 16.717(0.024) & 16.546(0.027)\\ 
2011-11-18 & 117.88 & 55883.28 & 18.191(0.043) & 17.644(0.036) & 16.832(0.032) & 16.710(0.027)\\ 
2011-11-30 & 129.83 & 55895.23 & 18.599(0.034) & 18.485(0.023) & 17.533(0.026) & 17.719(0.031)\\ 
\hline
\end{tabular}\label{tab:mlo}\end{minipage}
\end{center}
Note: All photometric observations were made with the Mount Laguna Observatory 40-inch telescope. \\
$^{a}$yyyy-mm-dd. \\
$^{b}$Days since first detection, 2011-07-22.9 UT (MJD 2,455,765.4).
\end{table*}

\subsection{Mount Laguna $BVRI$ Photometry}

All photometric data were collected using the Mount Laguna Observatory
(MLO; Smith \& Nelson 1969) 40-inch reflecting telescope, which is
equipped with a $2048 \times 2048$ pixel CCD camera (manufactured by
Fairchild Imaging Systems; CCD447) located at the $f/7.6$ Cassegrain
focus, providing a field of view of approximately $13.5\arcmin \times
13.5\arcmin$ with $0.41\arcsec$ pixel$^{-1}$.  The ``seeing,'' estimated 
from the full width at half-maximum intensity (FWHM) of stars on the CCD
frame, was generally $\sim 2\arcsec$, and exposure times of 5 to 30
minutes were typical for the observations, which were taken in the
standard Johnson-Cousins (Johnson et al. 1966; Cousins 1981) $BVRI$
bandpasses.

CCD frames were flatfielded using either twilight-sky or dome flats in
the usual manner, and cosmic rays were removed using the L.A. Cosmic
(van Dokkum 2001) algorithm.  Considerable fringing remained in $I$-band
images that did not properly flatten, likely produced by the varying
intensity of night-sky emission lines.  This has minimal impact on the
photometry since both the SN and the comparison stars were much
brighter than the background.

Figure~\ref{fig:finder} shows an MLO $B$-band image of NGC 151 taken
on 2011 August 5, with 10 ``local
standards'' identified in the field of PTF11iqb, which were used to
measure the relative SN brightness on nonphotometric nights.  The
absolute calibration of the field was accomplished on the photometric
night of 2012 November 19 by observing several fields of Landolt
(1992) standards over a range of airmasses.  The derived $BVRI$
magnitudes of the stars are given in Table~\ref{tab:stand}, along with
the estimated uncertainties.  The transformation coefficients to the
standard Johnson-Cousins systems were derived using the solutions from
this night.  We determined the instrumental magnitudes for the
standards using aperture photometry with the IRAF\footnote{IRAF is
  distributed by the National Optical Astronomy Observatories, which
  are operated by the Association of Universities for Research in
  Astronomy, Inc., under cooperative agreement with the National
  Science Foundation.} DAOPHOT package (Stetson 1987, 1991), which
yielded colour terms for the MLO observations of the form
\begin{eqnarray}
B & = & b + 0.093\,(B - V) + C_B,\nonumber\\
V & = & v - 0.067\,(B - V) + C_V,\\
R & = & r + 0.093\,(V - R) + C_R,\nonumber\\
I & = & i + 0.005\,(V - I) + C_I,\nonumber
\label{eqn:photometric_solutions}\nonumber
\end{eqnarray}
\noindent where $bvri$ are the instrumental and $BVRI$ the standard
Johnson-Cousins magnitudes. The terms $C_B$, $C_V$, $C_R$, and $C_I$ are
the differences between the zero-points of the instrumental and
standard magnitudes, determined for each observation by measuring the
offset between the instrumental and standard magnitudes and colours of
local standards.

We determined the instrumental magnitudes for PTF11iqb and the local
standards using the point-spread function (PSF) fitting technique within
DAOPHOT (Stetson 1991, and references therein).  We used only the
inner core of PTF11iqb and the local standards to fit the PSF in order
to reduce errors that can be introduced when there is a strong
gradient in the background (e.g., Schmidt et al. 1993).  In practice,
this core was generally set to be about the FWHM of a given image.
While the fitting radius of the SN and comparison stars was varied
from night to night to match the seeing, the sky background of the SN
and local standards was always set to an annulus with a radius of 20--30 
pixels ($8\farcs2$--$12\farcs3$) to maintain consistency
throughout the observations.  We next subtracted the mode of the sky
background\footnote{See Da Costa (1992) for a discussion of the
  advantages of using the mode of the background region rather than
  the mean or median.} to derive the instrumental magnitudes for the
SN and local standards. The transformation to the standard
Johnson-Cousins system was then accomplished using the colour terms
listed in Equation 1 and the standard magnitudes of the local
standards given in Table~\ref{tab:stand}.  On the two nights where
images in all four filters were not obtained, for the purpose of
estimating the appropriate colour term the colour of PTF11iqb was
estimated through interpolation or extrapolation from temporally
nearby values; since the colour terms are small, this approximation
should have little impact on the resulting photometry.

\begin{figure*}
\includegraphics[width=5.0in]{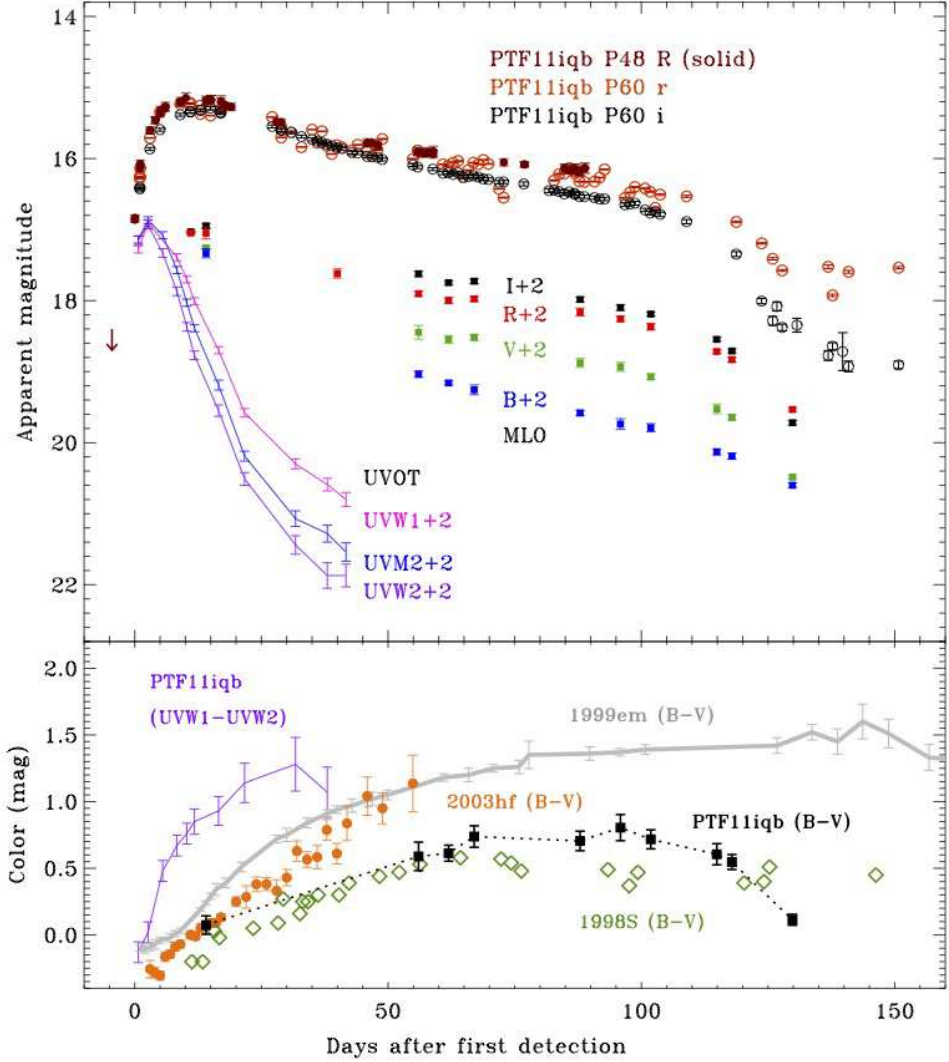}
\caption{{\it Top:} Apparent magnitudes for the early part of the
  light curve for PTF11iqb, with P48 $R$-band in solid red dots, and
  with P60 $r^{\prime}$ and $i^{\prime}$ bands shown with unfilled
  circles in red-orange and black, respectively (see Tables
  \ref{tab:p48} and \ref{tab:p60}).  {\it Swift}/UVOT ultraviolet
  photometry in the UW1, UM2, and UW2 filters is shown in magenta,
  blue, and purple, respectively (see Table~\ref{tab:uvot}).  UVOT
  magnitudes are offset by +2 for clarity.  $BVRI$ photometry from
  Mount Laguna Observatory (MLO) is shown in blue, green, red, and
  black squares (Table~\ref{tab:mlo}), with all four MLO bands offset
  by +2 mag for clarity.  These are all observed apparent magnitudes
  (i.e., not corrected for Milky Way reddening). {\it Bottom:} Colour
  evolution of PTF11iqb in the UV (UVW1--UVW2; purple) and visible
  wavelengths ($B-V$; black squares).  For comparison, $B-V$ colours
  for the SN II-P SN~1999em (thick grey line; \citealt{leonard02}), SN
  II-L 2003hf (orange circles; from \citealt{faran14}), and the SN~IIn
  SN~1998S (green diamonds; \citealt{fassia00}) are shown.}
\label{fig:photUVOT}
\end{figure*}

The final reported photometry was accomplished by taking the simple
mean of the values obtained using each of the available calibrator
stars on a given night; not all stars were available on all nights,
owing either to field-of-view limitations or saturation.  The results of
our photometric observations are given in Table~\ref{tab:mlo} and
shown in Figure~\ref{fig:photUVOT}.  The reported uncertainties come
from two sources.  First, there is the photometric uncertainty
reported by the error-analysis package in DAOPHOT from the statistics
of the SN and background region.  Second, there is uncertainty in the
transformation to the standard system.  We estimate the transformation
error by taking the standard deviation of the spread of the standard
magnitudes obtained using each of the local standard stars.  The
photometric and transformation errors were then added in quadrature to
obtain the uncertainty reported in Table~\ref{tab:mlo}; in nearly
every case, the total error was dominated by the uncertainty in the
transformation.

\begin{table*}\begin{center}\begin{minipage}{3.9in}
      \caption{Swift/UVOT photometry of PTF11iqb}
\scriptsize
\begin{tabular}{@{}lcccccc}\hline\hline
MJD     &UM2 (mag)     &$\sigma$ (mag) &UW1 (mag)  &$\sigma$ (mag) &UW2 (mag) &$\sigma$ (mag) \\ \hline
55766.1      &15.14    &0.05      &15.28    &0.05      &15.15    &0.06 \\ 
55768.0      &14.87    &0.05      &14.91    &0.05      &14.93    &0.06 \\
55770.9      &15.08    &0.05      &  ...        & ...         &15.33    &0.06 \\
55773.7      &15.57    &0.05      &15.39    &0.05      &15.87    &0.06 \\
55775.6      &16.03    &0.05      &15.70    &0.05      &16.37    &0.06 \\
55777.2      &16.39    &0.05      &16.01    &0.05      &16.77    &0.06 \\
55781.9      &17.19    &0.07      &16.70    &0.05      &17.55    &0.08 \\
55787.1      &18.19    &0.07      &17.58    &0.06      &18.51    &0.09 \\
55797.1      &19.07    &0.11      &18.30    &0.07      &19.44    &0.13 \\
55803.4      &19.28    &0.12      &18.59    &0.09      &19.87    &0.18 \\
55807.1      &19.54    &0.13      &18.80    &0.10      &19.87    &0.16 \\
\hline
\end{tabular}\label{tab:uvot}\end{minipage}
\end{center}
\end{table*}

\subsection{Swift/UVOT Ultraviolet Photometry}

The Ultra-Violet Optical Telescope (UVOT; \citealt{rkm+05}) onboard
\textit{Swift} observed PTF11iqb in the $UVW1$, $UVM2$, and $UVW2$
filters beginning on 2011 July 24.  We photometered the UVOT data
using standard procedures within the HEASoft\footnote{{\tt
    http://heasarc.nasa.gov/lheasoft/}} environment.  We used a
3\arcsec\ diameter aperture to extract flux from the transient, and
subtracted coincidence-loss-corrected count rates from underlying host
emission obtained from images in 2014 (e.g., \citealt{bhi+09}).
Photometric calibration was performed using the recipes from
\citet{pbp+08}.  The resulting magnitudes, all corrected to the AB
system, are displayed in Table~\ref{tab:uvot}.

\subsection{Spectroscopy}

\begin{figure}
\includegraphics[width=3.2in]{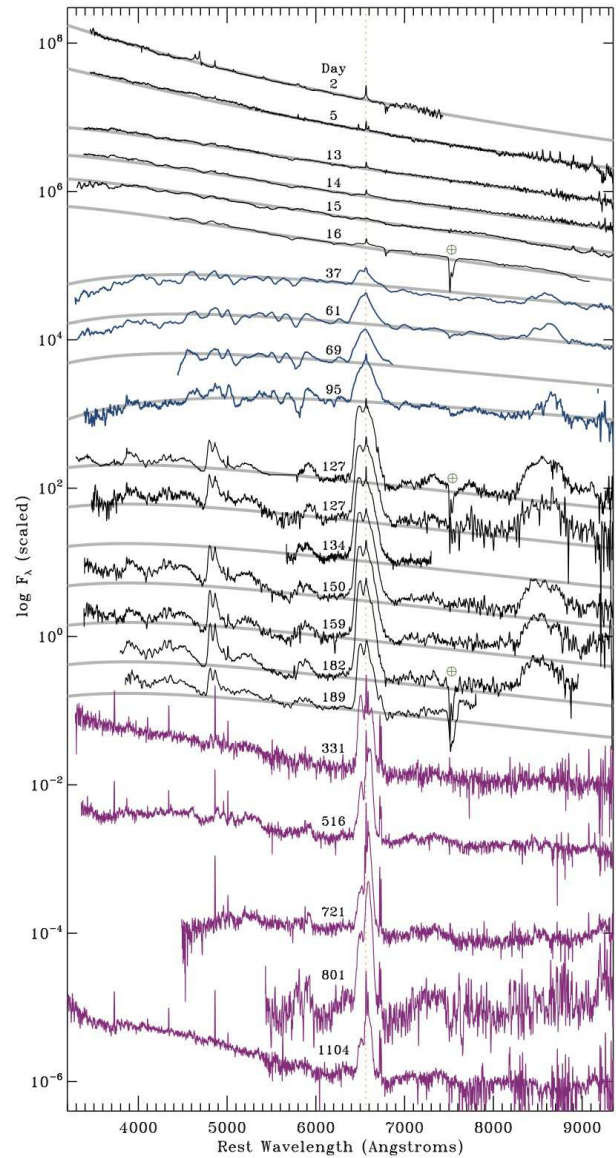}
\caption{Low/moderate resolution spectra of PTF11iqb (see
  Table~\ref{tab:spectab}), with early times at the top and later
  times at the bottom.  The four main phases discussed in the text are
  marked with changing colours: early times with CSM interaction
  (black), the plateau (blue), the early nebular phase (black), and
  the late nebular phase dominated again by CSM interaction (magenta).
  Strong telluric features are marked with ``{\earth}''.}
\label{fig:spec}
\end{figure}

PTF11iqb was brightening quickly at the time of discovery and was
presumably caught very early after explosion (see below), so we
quickly initiated spectroscopic follow-up observations starting on day
2 after first detection (1 day after discovery).  We obtained several
epochs of optical spectroscopy of PTF11iqb, which were densely sampled
at early times (see Fig.~\ref{fig:phot}, where epochs of spectroscopic
observations are plotted above the light curve).  We obtained spectra
of PTF11iqb using a number of different facilities, including the
Bluechannel (BC) spectrograph on the 6.5-m Multiple Mirror Telescope
(MMT), the Multi-Object Double Spectrograph (MODS; \citealt{bo00}) on
the LBT, the Double Beam Spectrograph (DBSP; \citealt{og82}) on the
Palomar 200-inch telescope (P200), the Intermediate dispersion
Spectrograph and Imaging System
(ISIS\footnote{http://www.ing.iac.es/PR/wht\_info/whtisis.html}) on
the 4.2-m William Herschel Telescope (WHT), GMOS on Gemini South, the
Kast spectrograph \citep{1993ms} on the Lick 3-m Shane reflector, the
Low-Resolution Imaging Spectrometer (LRIS; \citealt{oke95}) mounted on
the 10-m Keck~I telescope, and the Deep Imaging Multi-Object
Spectrograph (DEIMOS; \citealt{faber03}) on Keck~II.  Details of the
spectral observations are summarised in Table~\ref{tab:spectab}, and
the spectra will be released on the Weizmann Interactive Supernova
Data Repository (WISeREP; \citealt{yaron12}; {\tt
  http://www.weizmann.ac.il/astrophysics/wiserep/}).  The slit was
always oriented at the parallactic angle \citep{filippenko82}, and the
long-slit spectra were reduced using standard procedures.  Final
spectra are shown in Figure~\ref{fig:spec}, although a few epochs of
spectra are not displayed in this figure because they were taken close
in time to another spectrum that appears very similar.  The earlier
epochs in Figure~\ref{fig:spec} have a blackbody plotted in grey for
each observed spectrum; these correspond to the temperatures listed as
$T_{\rm BB}$ in Table~\ref{tab:spectab}, although these are intended
only as a rough relative comparison because they depend on the adopted
reddening.  Nevertheless, inferred temperatures around 7000~K in the
plateau phase suggest that these quoted temperatures are not wildly in
error.  Details of the H$\alpha$ line profile are shown in
Figure~\ref{fig:halpha}.

\begin{table}\begin{center}\begin{minipage}{3.25in}
      \caption{Spectroscopic Observations of PTF11iqb}
\scriptsize
\begin{tabular}{@{}llcccc}\hline\hline
Date     &Tel./Instr &Day &$\Delta\lambda$(\AA) & &$T_{\rm BB}$(K) \\ \hline
2011\,Jul\,24 &GS/GMOS      &2   &3500--7400  &  &25,000   \\
2011\,Jul\,24 &P200/DBSP    &2   &3520--9825  &  &25,000   \\
2011\,Jul\,27 &WHT/ISIS     &5   &3500--9500  &  &16,000   \\
2011\,Jul\,30 &WHT/ISIS     &8   &3500--9500  &  &...   \\
2011\,Aug\,01 &Keck2/DEIMOS &10  &5150--7800  &  &...   \\
2011\,Aug\,04 &Lick/Kast    &13  &3436--9920  &  &10,500   \\
2011\,Aug\,05 &Lick/Kast    &14  &3436--9920  &  &10,500   \\
2011\,Aug\,06 &P200/DBSP    &15  &3520--9825  &  &10,000   \\
2011\,Aug\,07 &WHT/ISIS     &16  &3500--9500  &  &10,000   \\
2011\,Aug\,28 &P200/DBSP    &37  &3520--9825  &  &6,500   \\
2011\,Sep\,21 &P200/DBSP    &61  &3520--9825  &  &6,500   \\
2011\,Sep\,26 &Keck1/LRIS   &66  &3200--7350  &  &...   \\
2011\,Sep\,29 &Keck2/DEIMOS &69  &4500--9300  &  &6,500   \\
2011\,Oct\,25 &Lick/Kast    &95  &3436--9920  &  &5,600   \\
2011\,Nov\,26 &Keck1/LRIS   &127 &3340--10300 &  &7,500   \\
2011\,Nov\,26 &Lick/Kast    &127 &3436--9920  &  &7,500   \\
2011\,Dec\,02 &Keck1/LRIS   &134 &3300--7400  &  &7,500   \\
2011\,Dec\,18 &Lick/Kast    &150 &3436--9920  &  &7,500   \\
2011\,Dec\,27 &Lick/Kast    &159 &3436--9920  &  &7,500   \\
2012\,Jan\,02 &MMT/BC       &165 &5550--7500  &  &...   \\
2012\,Jan\,19 &MMT/BC       &182 &3820--9000  &  &7,500   \\
2012\,Jan\,26 &Bok/SPOL     &189 &3900--7600  &  &7,500   \\
2012\,Jan\,30 &MMT/BC       &193 &5550--7500  &  &...   \\
2012\,Jun\,16 &Keck1/LRIS   &331 &3330--10100 &  &...   \\
2012\,Jun\,28 &MMT/BC       &343 &5550--7500  &  &...   \\
2012\,Dec\,18 &Keck1/LRIS   &516 &3400--10200 &  &...   \\
2013\,Jan\,14 &MMT/BC       &543 &5550--7500  &  &...   \\
2013\,Jul\,11 &Keck2/DEIMOS &721 &4400--9600  &  &...   \\
2013\,Sep\,04 &MMT/BC       &775 &nondetect.  &  &...   \\
2013\,Sep\,29 &LBT/MODS     &801 &5500--10000 &  &...   \\
2014\,Jul\,29 &Keck/LRIS   &1104 &3400--10200 &  &...   \\
\hline
\end{tabular}\label{tab:spectab}\end{minipage}
\end{center}
\end{table}

Since our P48 and P60 photometry covers early times well, but includes
poor sampling after day 150 (and only upper limits in the $i^{\prime}$
band on days 500--600), we supplement our imaging photometry by using
our flux-calibrated spectra.  In spectra taken after day 150, we
measured integrated fluxes within specific wavelength ranges to
simulate the flux that would be observed in the $R$ (5700--7300 \AA),
$r^{\prime}$ (5500--6700 \AA), and $i^{\prime}$ (6900--8100 \AA)
filters, and then converted these fluxes to magnitudes and added them
to Figure~\ref{fig:phot} as solid rectangles with the same symbol
colours as the imaging photometry.  The uncertainty here is difficult
to quantify, since the main source of error is the exact positioning
of the target and associated standard star within the slit, but we
adopt representative errors of $\pm0.4$ mag in Figure~\ref{fig:phot}
(at earlier times when our photometry overlaps with spectra, the
spectrophotometry generally agree to 0.4 magor better).  As discussed
in more detail below, it is important to recognise that at the
redshift of PTF11iqb, the main difference between the $R$ and
$r^{\prime}$ filters is that $R$ includes the full H$\alpha$
emission-line profile, whereas the red end of the $r^{\prime}$ filter
cuts off the red wing of PTF11iqb's H$\alpha$ line.  It is also
important to note that the pseudo-photometry from spectra subtracts
the background by sampling host-galaxy light on either side of the SN,
whereas the imaging photometry utilises background subtraction with a
SN-free reference image.  Thus, if there is an underlying star cluster
or other coincident source, it will be included in the spectral
photometry but will be absent in the imaging photometry.  This may be
important at late times when PTF11iqb is faint, so the true brightness
of PTF11iqb may be below the level indicated by the spectrophotometry.
Nevertheless, these measurements provide a useful guide for the
late-time evolution.

\section{RESULTS}

\subsection{Light Curve}

The multiband light curves and colour evolution for the early phases
of PTF11iqb are shown in Figure~\ref{fig:photUVOT}, while
Figure~\ref{fig:phot} shows a subset of the data on an
absolute-magnitude scale and extended to later times, as compared to a
few previously observed SNe.  The light curve displays a rapid rise in
only $\sim10$ days to a peak luminosity of about $-18.4$ mag.  This is
followed by a decline that flattens out in a plateau or shoulder at
about $-17$ mag, dropping off the plateau after 100--120 days.

To the extent that flux proportional to $t^2$ is a valid description
of the early rise, the dotted curve in Figure~\ref{fig:phot} suggests
that explosion occurred at roughly $-3.3$ days (i.e., 3.3 days before
day 0, defined here as the first detection on 2011 July 22.37).  This
is consistent with the upper limit of 18.5 mag (unfiltered) for our
nondetection on 2011 July 17.47 (Parrent et al.\ 2011), indicated by
the red arrow in Figure~\ref{fig:phot}.  Since this upper limit
corresponds to an absolute magnitude of around $-15$ mag, it does not
place strong constraints on the luminosity or duration of a possible
pre-SN eruption.  The relatively quick rise to peak in only a few days
seems to suggest an extended progenitor star similar to a RSG, which
is a topic we discuss in more detail later.

The visible-wavelength colour evolution of PTF11iqb (see
Figure~\ref{fig:photUVOT}) starts out very blue and becomes steadily
redder for the first 60--70 days, and then seems to level off at $B-V
\approx0.75$ mag before becoming more blue again at late times.
PTF11iqb never gets as red as a normal SN~II-P, like SN~1999em,
probably owing to the continued influence of CSM interaction.  The
$B-V$ colour evolution of PTF11iqb is quite similar to that of
SN~1998S, although at early times it is also similar to a Type II-L
event like SN~2003hf (the photometry here is from the recent study of
SNe~II-L by \ \citealt{faran14}).  We demonstrate later that the
spectra around days 30--60 most closely resemble SNe II-L at similar
times.  The UV colours are very blue at early times, but quickly
redden as the UV luminosity plummets and the photosphere cools rapidly
in the first 30--40 days after explosion.

There is little change in colour or luminosity during the latter part
of the plateau phase from days $\sim50$ to 120.  After the drop from
this plateau, PTF11iqb exhibits a nebular phase that (at least in the
$i^{\prime}$ band) roughly follows the luminosity decline expected for
radioactive decay from $^{56}$Co (see Figure~\ref{fig:phot}).  While
PTF11iqb is qualitatively similar to the light-curve behaviour of
SNe~IIn-P in this respect \citep{mauerhan13b,smith13}, the drop after
the plateau was not nearly as severe, and the late-time radioactive-decay 
luminosity was not as low.  Also, as we discuss below, during
the decline from peak and the plateau phase, the spectrum does not
show a strong Type~IIn signature (resembling SNe~1994W and 2011ht),
but more closely resembles spectra of SNe~II-P and II-L.  Thus, its
spectral evolution was unlike the rather homogeneous class of SNe~IIn-P 
\citep{mauerhan13b}, which exhibit a persistent SN~IIn spectrum
throughout their evolution.  For these reasons we do not classify
PTF11iqb as a Type~IIn-P event, although we acknowledge that SN
classification criteria may be inadequate if an object morphs through
different types as it evolves.

Based on an analysis of the spectra (below), it is clear that the
$i^{\prime}$ band traces the true continuum luminosity much better
than the $R$ or $r^{\prime}$ bands.  After the drop of the plateau in
the light curve at 100--120 days, the red filters are dominated by
strong H$\alpha$ emission from CSM interaction, to a different degree
in each filter because of their different wavelength responses (see
above).  The $i^{\prime}$ band does not include the bright H$\alpha$
line.  From the time of the plateau drop until about day 500, the
$i^{\prime}$ band is consistent with radioactive decay from $^{56}$Co,
and in fact has about the same luminosity and decline rate as the
prototypical Type II-P SN~1999em.  This would imply a synthesised
$^{56}$Ni mass of roughly 0.03\,M$_{\odot}$ (for SN~1999em, this was
estimated as 0.02\,M$_{\odot}$ by \citealt{elmhamdi} and
0.036\,M$_{\odot}$ by \citealt{utrobin}), and hence, a normal
core-collapse SN~II-P.

\begin{figure*}
\includegraphics[width=6.0in]{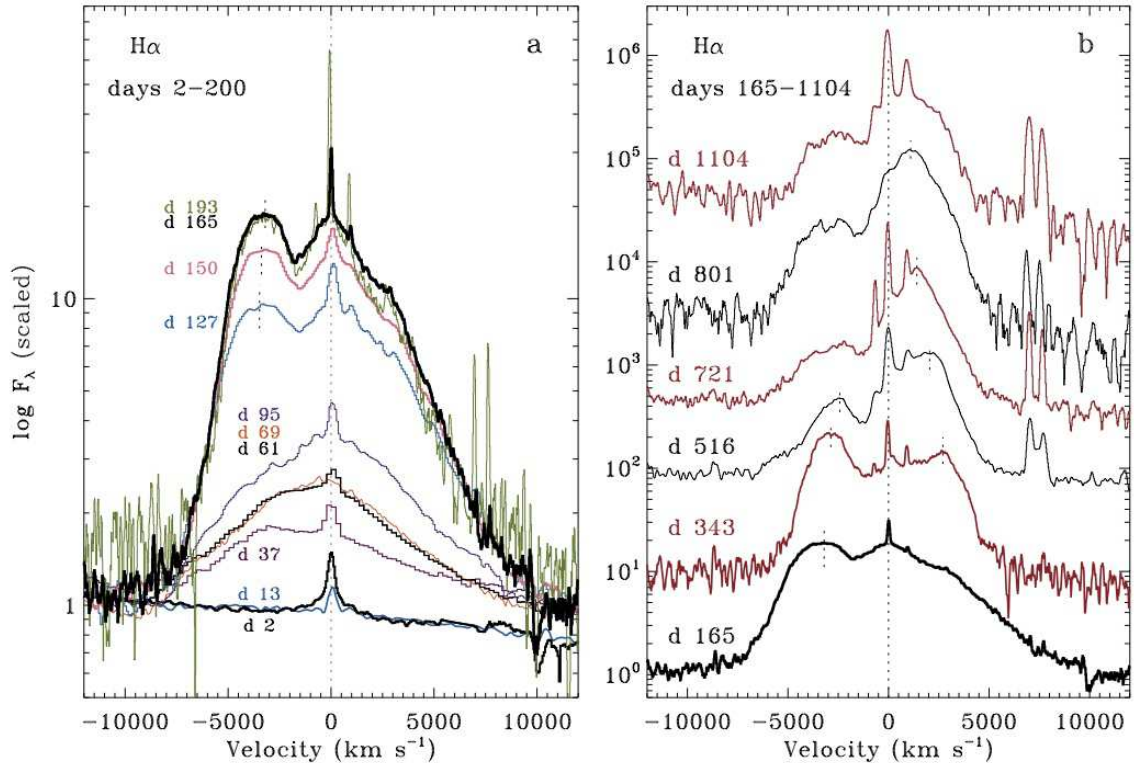}
\caption{Details of the H$\alpha$ line profile of PTF11iqb.  (a) The
  left panel shows relatively early-time data, from the time of
  discovery until the beginning of the nebular phase.  These spectra
  are normalised to the continuum level, and the time sequence of line
  strength (including broad components) leads to spectra that
  generally go up with time on this plot.  (b) The right panel shows
  late-time spectra during the nebular phase from days 165 onward.
  Short vertical lines (dotted) mark velocities measured for
  relatively well-defined red and blue peaks (see text and
  Figure~\ref{fig:model}).}
\label{fig:halpha}
\end{figure*}

As discussed later, we attribute the excess luminosity (compared to
SN~1999em) in the first half of the plateau to CSM interaction, as
well as the strong late-time H$\alpha$ emission.  At the very latest
phases (after day 500) the true underlying continuum level is very
uncertain, since photometry from template-subtracted images only
provides upper limits, whereas the pseudo-photometry estimated from
spectra does not correct for a possible underlying star cluster or
other faint coincident source, and may therefore be an overestimate of
PTF11iqb's luminosity by an amount that is difficult to quantify.

\subsection{Spectral Evolution}

Here we provide a brief overview of the main stages in the spectral
evolution of PTF11iqb.  The spectra are discussed in more detail below
in the comparison with SN~1998S, as well as in sections discussing the
late-time H$\alpha$ and early-time WR features specifically. We
highlight three main stages: (1) at very early times around peak
luminosity, (2) days 20--100, corresponding to the latter part of the
plateau in the light curve, and (3) during the nebular phase after day
$\sim120$.

\subsubsection{Early Spectral Evolution}

For the first $\la20$ days after explosion, the spectrum of PTF11iqb
was characterised by a smooth blue continuum matched by a blackbody
that cools quickly from about $\gtrsim$25,000~K on day 2 to about
10,000 K on day 16.  The spectrum on day 2 shows a narrow (unresolved
at FWHM $\approx$ 500 km s$^{-1}$ at this epoch) H$\alpha$ line core
with broader and symmetric Lorentzian wings extending to about
$\pm2000$ km s$^{-1}$, and also shows a comparably strong
``Wolf-Rayet'' emission feature (a combination of He~{\sc ii}
$\lambda$4686 and C~{\sc iii}/N~{\sc iii}).  By day 5 these features
have mostly faded; thereafter, only narrow Balmer lines remain atop
the smooth blue continuum.  Similar WR features have been seen in a
few other SNe at early times; this is discussed more in \S 7.

\subsubsection{Plateau Spectral Evolution}

Starting on day 16, we begin to see an underlying broad component of
H$\alpha$ that continues to grow in strength as the narrow component
fades (Figures~\ref{fig:spec} and \ref{fig:halpha}a).  This broad
component arises in the underlying fast SN ejecta.  There is some
persistent very faint narrow H$\alpha$ emission during the plateau
phase, but it is difficult to ascertain how much of it may be
contributed by underlying H~{\sc ii} regions, since its strength
varies from one spectral epoch to the next (which could be caused by
true intensity changes, or by various amounts of surrounding H~{\sc
  ii} region emission being included in the slit).  In any case, the
narrow component of H$\alpha$ is faint (a few percent of the total
H$\alpha$ flux).  On the other hand, the radius of the CSM interaction
front in our model (see below) and the SN ejecta photospheric radius
are comparable at this epoch, so perhaps the CSM interaction intensity
remains steady, but much of the shock interaction region is overtaken
by the opaque SN ejecta photosphere (see below).

\begin{figure}
\includegraphics[width=3.1in]{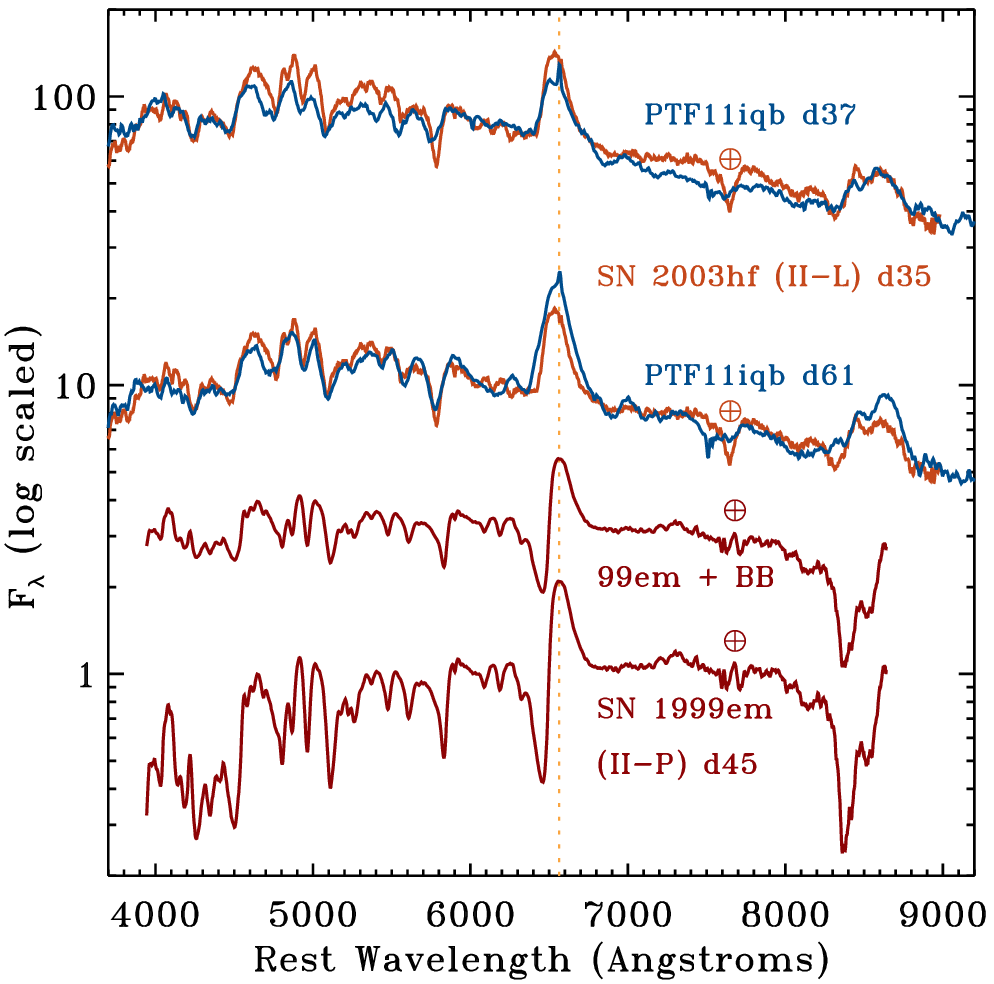}
\caption{Comparison of visible-wavelength spectra of PTF11iqb on days
  37 and 61 (blue) with those of SNe~II-L (SN~2003hf; orange)
  and II-P (SN~1999em; red). For SN~1999em, we show the observed day
  45 spectrum, as well as that same spectrum added to a blackbody to
  dilute the strength of absorption features.  These spectra of
  SNe~2003hf and 1999em were obtained from the UC Berkeley SN database,
  and appeared in \citet{faran14} and \citet{leonard02},
  respectively.}
\label{fig:comp}
\end{figure}

After about a month, the featureless hot blue continuum from the
earlier phase transforms into a spectrum dominated by broad absorption
and emission profiles, as in normal ejecta-dominated SNe like SNe II-L
and II-P.  Besides the broad H$\alpha$ line with a width of
$\sim$\,10,000 km s$^{-1}$ (FWZI), we see numerous absorption features
in the blue typical of SNe~II-P \citep{dessart13}, as well as the
broad Ca~{\sc ii} near-infrared (IR) triplet in emission.  Some of the
absorption (especially H$\alpha$ and Ca~{\sc ii}) appears muted
compared to canonical SN~II-P atmospheres.  Indeed, PTF11iqb's
relatively weak P-Cygni absorption in H$\alpha$ and Ca~{\sc ii} more
closely resembles that of SNe~II-L than SNe~II-P, while at 4000--6000
\AA, SNe~II-P and II-L look qualitatively very similar (see
Figure~\ref{fig:comp}). \citet{gut14} discuss the diversity in the
ratio of H$\alpha$ absorption to emission ($a/e$) in SNe II-P and
II-L, and PTF11iqb resembles objects in that study with lower $a/e$
vaues.  So overall, even though the light curve shows a clear
underlying plateau, the spectrum during the plateau phase looks more
like that of SNe~II-L; we suspect that this may be due to reionisation
of the outer SN ejecta by CSM interaction.  As we discuss below, the
SN~II-L-like spectral appearance is shared by SN~1998S at similar
times in its evolution.

\subsubsection{Late-Time Spectral Evolution}

After the drop in luminosity at the end of the plateau (after days
100--120), the H$\alpha$ emission from PTF11iqb starts to become quite
interesting, exhibiting an asymmetric and multipeaked profile
(Figure~\ref{fig:halpha}b).  During this stage, H$\alpha$ develops a
strong and somewhat broad blueshifted emission peak centered at
$-$3000 to $-$3500 km s$^{-1}$.  The red side of the line shows no
similar peak, but instead begins to drop off sharply at around +3000
km s$^{-1}$.  This asymmetric blueshifted H$\alpha$ profile changes
little over the time period from days 120 to 200
(Figure~\ref{fig:halpha}a).

Blaming this strong blueshifted asymmetry of H$\alpha$ on dust or
opaque SN ejecta (blocking the receding side) would seem appealing,
were it not for the fact that the asymmetric H$\alpha$ profile then
transitions from a blueshifted asymmetry to a much stronger {\it
  redshifted} peak between days 331 and 516
(Figure~\ref{fig:halpha}b).  In the day 516 spectrum and afterward,
the red peak of H$\alpha$ remains stronger than the blue peak (even if
dust that was previously blocking the red side were destroyed, it
would not make the red side brighter; true asymmetry in the CSM
density is needed).  We note that a delay in light-crossing time from
the back to the front of the SN cannot explain this shift from blue to
red peaks, since the light-crossing time is less than 10 days at radii
of $\sim500$ AU.

Interestingly, the strength of the blueshifted peak coincides
temporally with a lingering blue continuum and broad SN ejecta lines
like the Ca~{\sc ii} IR triplet.  These are strong in the early phases
of the nebular decay, as in typical SNe II-P.  From day 516 onward,
however, the blueshifted peak of H$\alpha$, the blue continuum, and
the broad ejecta lines all fade in tandem.  When the red peak is
strongest (after day 500), the overall spectrum is dominated by
H$\alpha$ with very little continuum contributing to the observed
spectrum.  This probably signifies that the spectrum at days 120--200
has a substantial contribution from the normal radioactivity-powered
nebular emission from the inner ejecta, and this is when the $i$-band
light curve and the H$\alpha$ line luminosity (see below) both follow
the $^{56}$Co decay rate.  At the latest times (after 300--500 days),
the spectrum is dominated by CSM interaction because the radioactivity
has faded much faster than the ongoing CSM interaction luminosity.

A narrow emission component of H$\alpha$ is seen at several epochs,
most likely tracing the preshock CSM.  Our highest-resolution spectra
are those epochs obtained at late times with the MMT using the 1200
line mm$^{-1}$ grating, which are most suitable to constrain the
preshock wind speed. Examining epochs that appear to have less
contamination from surrounding H~{\sc ii} regions (days 37, 61, 69,
95, 127, 150, 165; probably owing to better seeing, evidenced by
weaker [N~{\sc ii}] emission adjacent to H$\alpha$), we find that the
narrow component of H$\alpha$ has Gaussian FWHM values of $\sim80$ km
s$^{-1}$, comparable to the spectral resolution of the data.  The
preshock wind is therefore much slower than that of a typical WR star
($\sim10^3$ km s$^{-1}$ or more), or that of a blue supergiant or
luminous blue variable (LBV; 100--500 km s$^{-1}$), and closer to
values seen in cool red or yellow supergiants (YSGs; see
\citealt{smith14}).  We quote 100 km s$^{-1}$ as a fiducial value for
the CSM expansion speed below, but the actual wind speed and mass-loss
rates may be smaller by a factor of 2--3.

\begin{figure}
\includegraphics[width=3.2in]{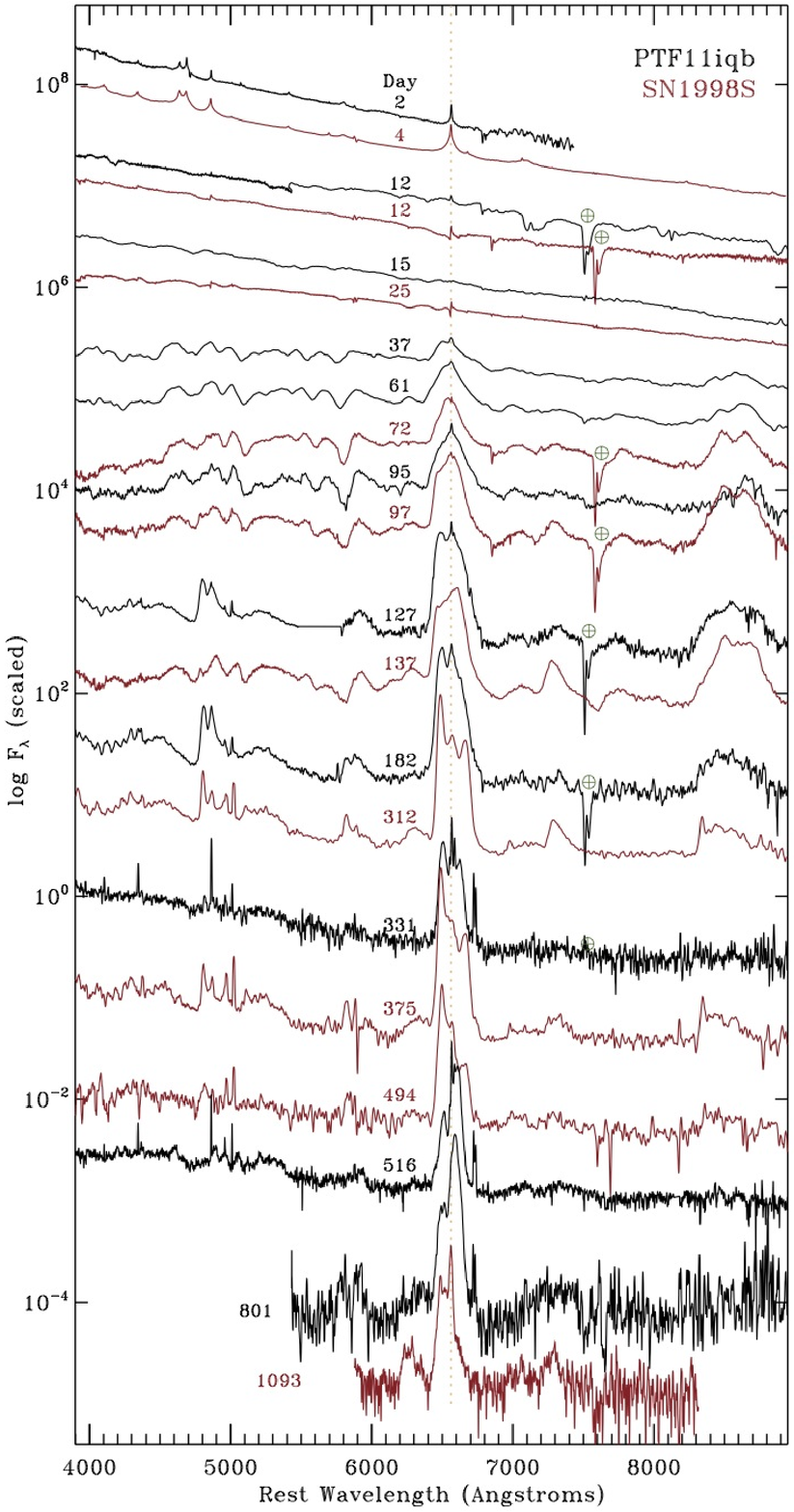}
\caption{The spectral evolution of PTF11iqb compared to that of
  SN~1998S. All spectra of PTF11iqb are black, and all spectra of
  SN~1998S are red. Spectra of PTF11iqb for various epochs are the
  same as in Figure~\ref{fig:spec}.  Spectra of SN~1998S are from
  \citet{leonard00} (days 4, 25, 137, 312, 375, and 494),
  \citet{fassia01} (days 12, 72, and 97), and \citet{pozzo04} (day
  1093).  The days for SN~1998S are as listed by those authors; this
  could cause offsets of a few days compared to PTF11iqb, which may be
  relevant at early times when comparing the two SNe, but which will
  not be significant at later times.}
\label{fig:spec98s}
\end{figure}
\section{A Comparison with SN~1998S}

Readers familiar with SNe~IIn may recognise that our description of
the spectral evolution of PTF11iqb in \S 3.2 sounds remarkably similar
to that of the classic well-observed Type IIn object SN~1998S.
Figure~\ref{fig:spec98s} compares the spectral evolution of the two
SNe, with visual-wavelength spectra interleaved in chronological order
after explosion.  The spectra of SN~1998S from \citet{leonard00} are
taken from the UC Berkeley SN database \citep{silverman12}, and the
others are downloaded from the WISeREP \citep{yaron12} database,
originating from papers by \citet{fassia01} and \citet{pozzo04}.  The
days listed in Figure~\ref{fig:spec98s} are relative to discovery as
noted by those authors; the explosion date of SN~1998S is not known
precisely, but may be a few days before discovery \citep{leonard00}.

Basically, the spectra of the two SNe in Figure~\ref{fig:spec98s} are
almost identical in their evolution, sufficient to claim that PTF11iqb
is a near twin of SN~1998S.  Both SNe proceed through all the same
spectral changes outlined above at roughly similar times.  We note two
exceptions to this twinhood.

First, spectral signatures of CSM interaction are generally stronger
in SN~1998S at early times, consistent with its significantly higher
peak luminosity (Figure~\ref{fig:phot}).  In its day 2 spectrum,
PTF11iqb shows an almost identical broad-winged Lorentzian H$\alpha$
profile and Wolf-Rayet (WR) features to those that were first
described in detail for SN~1998S by \citet{leonard00}.
\citet{chugai01} noted how these can arise from CSM interaction in the
opaque inner wind.  An alternative explanation for the early-time
spectrum may be ionisation of a dense inner wind by a UV flash from
shock breakout, as was hypothesised to occur in SN~2013cu
\citep{galyam14}.  It is difficult to distinguish between these two
hypotheses with the available data, but in either case, PTF11iqb
requires an opaque inner wind at a radius of $\sim10$ AU.  An argument
in favour of CSM interaction being important is that the wind density
parameter required for the inner wind is very similar to that in the
outer wind required by the late-time CSM interaction when the emission
is optically thin (see \S 7).  In any case, the key difference as
compared to SN~1998S is that in PTF11iqb these features are weaker and
they fade more quickly.  In the day 2 spectrum of PTF11iqb, we measure
EW(H$\alpha$) $= 13.5\pm0.8$ \AA \ and EW(WR) $= 29.6\pm1.0$ \AA \
(emission-line EWs are positive).  The WR bump and strong Lorentzian
wings of H$\alpha$ are only seen in the day 2 spectrum and they fade
completely by day 5 in PTF11iqb, leaving only a narrow H$\alpha$
component.  In SN~1998S, by contrast, these features are still quite
strong in the days 3, 4, and 5 spectra shown by \citet{leonard00}.  In
the day 4 spectrum of SN~1998S in Figure~\ref{fig:spec98s}, we measure
EW(H$\alpha$) $=55\pm1.0$ \AA \ and EW(WR) $=39\pm1.0$ \AA.  Note that
because of the higher luminosity of SN~1998S, the line luminosities
are about an order of magnitude larger in SN~1998S.  It is also
interesting to note that the EW(WR)/EW(H$\alpha$) ratio is higher in
PTF11iqb, perhaps reflecting the fact that the spectrum was obtained
sooner after explosion than for SN~1998S, or perhaps higher ionization
at lower densities.

Second, PTF11iqb shows an interesting and significant difference in
its late-time H$\alpha$ profile evolution.  Both objects exhibit
qualitatively similar asymmetric and multipeaked H$\alpha$ lines at
late times.  However, PTF11iqb shifts from having a prominent blue
peak at days 120--200 to a very strong red peak after day 500.
SN~1998S does not do this, and stays with a blue peak continually.
This is significant because the persistent blue peak of SN~1998S at
late times was attributed to dust formation blocking the receding
parts of the system \citep{leonard00,pozzo04}.  The nearly identical
temporal evolution of PTF11iqb, but with the difference of ending with
a stronger redshifted peak, implies that this asymmetry may be caused
by nonaxisymmetric structure and viewing orientation, rather than
preferential extinction of the far side by dust.  The asymmetry in the
CSM is discussed more below (see \S 6).

Aside from these two points, the spectra of PTF11iqb and SN~1998S are
basically the same.  How does this fit with their light curves?
SN~1998S was roughly a factor of 4 more luminous at peak, consistent
with stronger CSM interaction, whereas the two had very similar
luminosities from about day 80 onward.  In the next section, we show
that we can approximate the light curve of PTF11iqb by adding the
luminosity from weak\footnote{Here we intend ``weak'' to mean that CSM
  interaction is weak enough that it does not dominate the visible
  wavelength luminosity of the SN ejecta photosphere in these early
  phases.} CSM interaction to the light curve of a normal SN~II-P.
This veils the sharp drop in a normal plateau, making it a more
gradual ``shoulder'' in the light curve.  If we crank up the
early-time CSM interaction in this scenario, the peak luminosity will
be higher, and the shoulder in the light curve owing to the underlying
SN~II-P plateau will be harder to recognise, resembling SN~1998S.
Thus, the inference that PTF11iqb was like SN~1998S but with weaker
CSM interaction seems qualitatively consistent with both the spectra
and light curves.

\begin{figure*}
\includegraphics[width=4.6in]{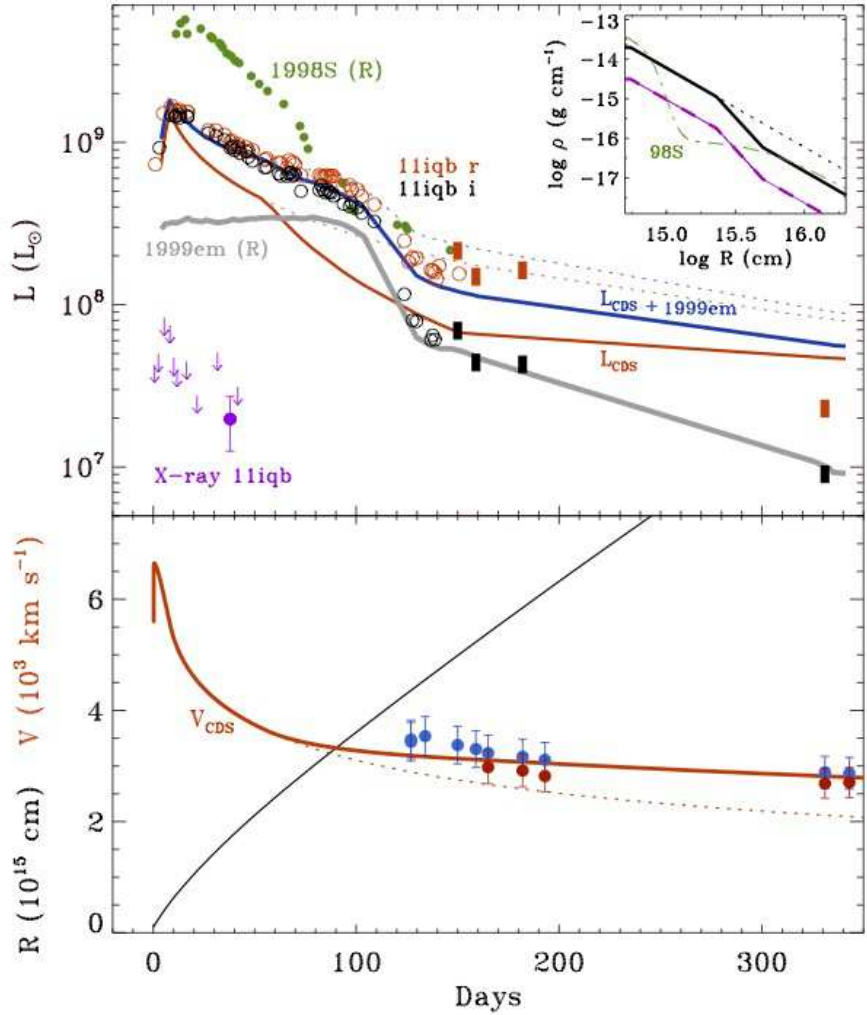}
\caption{A simple model for the light curve of PTF11iqb.  {\it Top
    panel:} Luminosities based on the observed $r$ and $i$-band light
  curves (red and black data, respectively) of PTF11iqb, compared to a
  simple model.  The thick blue curve that approximates the
  light-curve shape (labeled $L_{\rm CDS}$+1999em) is a combination of
  a normal SN~II-P light curve (SN~1999em, shown in grey; from
  \citealt{leonard02}) and a simple CSM interaction model (orange
  curve labeled $L_{\rm CDS}$).  This model results from the radial
  density distribution shown in the inset at upper right (solid
  black), but where we have assumed an efficiency of 15\% to account
  for aspherical geometry (see text for details).  The dashed purple
  curve in this panel is the equivalent density distribution with the
  same mass if the model were spherically symmetric. The dashed green
  curve is the radial density distribution from the model for SN~1998S
  by \citet{chugai01}.  In the luminosity plot (top), the dotted blue
  and dotted red/orange curves show what the total luminosity would
  look like with a constant wind density parameter $w$, corresponding
  to the dotted black line in the radial density plot.  (The $R$-band
  light curve of SN~1998S from \citet{fassia00} is shown for
  comparison, as are X-ray data for PTF11iqb from \citet{ofek13b}).
  {\it Bottom panel:} The evolution of the cold dense shell (CDS)
  radius (black) and velocity (rust coloured) with time in the CSM
  interaction model from the top panel.  The solid curve labeled
  $V_{\rm CDS}$ is the velocity for the solid black density
  distribution in the upper-right inset, and the dotted orange curve
  corresponds to constant $w$ for comparison.  The blue and red data
  are velocities of the blue and red peaks of the asymmetric H$\alpha$
  line (see Figure~\ref{fig:halpha}). }
\label{fig:model}
\end{figure*}

\section{A Model Light Curve}

In Figure~\ref{fig:model} we demonstrate that one can approximately
match the light curve of PTF11iqb using a SN~II-P light curve that has
extra luminosity from CSM interaction added to it.  The SN~II-P light
curve we adopt is that of SN~1999em \citep{leonard02}, but of course
this is just a convenient reference.  (We could have used a different
SN~II-P with a somewhat different light curve, and then adjusted the
CSM parameters accordingly.)  To this we add luminosity from CSM
interaction ($L_{\rm CDS}$; the luminosity of the cold dense shell),
calculated from a simple model with fast SN ejecta overtaking slower
dense CSM, and the loss in kinetic energy is converted to radiation
(see, e.g., \citealt{smith13a,smith13}).  Models of this type are not
unique and have degeneracy in adjustable parameters like explosion
energy, CSM mass, CSM radial distribution, and geometry.  The model
shown here is meant as a plausibility argument that PTF11iqb can be
explained with a normal core-collapse SN that has weaker CSM
interaction than SN~1998S, and provides only a very rough estimate of
the CSM conditions.

Observations of the CDS velocity at late times when the emission is
optically thin can help restrict some of the degeneracy in the model.
Velocities of the blue and red peaks observed in the late-time
H$\alpha$ profiles are plotted in the bottom panel of
Figure~\ref{fig:model}.  Normally, decelerating the fast $10^{51}$ erg
SN ejecta to the observed coasting speed of around 3000 km s$^{-1}$
would require a high CSM mass, and also that roughly half the
explosion kinetic energy was radiated away --- this would produce a
value of $L_{\rm CDS}$ that is much higher than observed in PTF11iqb.
To make a CSM-interaction model consistent with the observed
velocities without overproducing the luminosity, we must adjust the
emergent luminosity by an artificial efficiency factor of $\sim15$\%.
One option to accomplish this physically is to simply lower the CSM
density and the SN ejecta mass (and SN energy) to 15\%.  The lower SN
ejecta mass and energy would, however, then seem inconsistent with the
underlying plateau light curve and radioactive decay tail that
indicate a normal SN~II-P.  A second and simpler option is to
interpret the 15\% efficiency as a geometric effect --- i.e., if the
dense CSM only intercepts 15\% of the solid angle seen by the SN
ejecta, one can still have a normal $10^{51}$ erg explosion and
high-density CSM, but the resulting $L_{\rm CDS}$ is much lower than
it would be in a spherically symmetric model with the same parameters.
The 15\% geometric covering factor could, in principle, arise from CSM
that is in a disc or torus, a nonaxisymmetric shell, or clumps that
intercept an equivalent fraction of the solid angle.

Agreement between our model and observations does not provide
conclusive evidence of asymmetric CSM, but there are a number of
reasons why nonspherically symmetric CSM is plausible.  First, there
is clear observational precedent: the similar explosion SN~1998S was
inferred to have a highly aspherical, perhaps disc-like or toroidal
CSM based on spectropolarimetry \citep{leonard00}.  SN~1998S was also
inferred to have a disc-like CSM based on the multipeaked asymmetric
H$\alpha$ line profiles seen at late times
\citep{leonard00,gerardy00,fassia01,pozzo04}, which are very similar
to those of PTF11iqb.  A flattened CSM geometry was also inferred for
the more recent event SN~2009ip based on energetic arguments,
spectropolarimetry, and spectral clues
\citep{smith+14,mauerhan14,levesque14,ofek13c}.  Second, our spectra
do not exhibit the narrow P-Cygni absorption that one might expect if
the required preshock CSM density is along our line of sight to the SN
photosphere.  Last, evolved massive stars commonly show highly
aspherical CSM when their nebulae are spatially resolved (this is
discussed more in the next section).

A substantial fraction of $L_{\rm CDS}$ in our model should escape as
X-rays rather than being converted to visible-wavelength continuum or
H$\alpha$ emission.  Although the observed X-ray luminosity at early
times shows that only a small fraction of the X-ray luminosity escapes
(Figure~\ref{fig:model} and \citealt{ofek13b}), there have not yet
been published constraints for late-time X-ray emission from PTF11iqb.
This is expected theoretically for most SNe IIn
\citep{ci12,svirski12}.  From our analysis of the late-time H$\alpha$
(see below) we infer that the wind density parameter in the outer wind
may actually be similar to that in the very inner wind, arguing that
the drop in density at $10^{15.5}$ cm used to model the light curve
merely reflects a decreasing optical depth of the outer wind and
larger escape fraction of X-rays.  As the material expands to larger
radii and the optical depth drops, we would indeed expect a larger
fraction of the total postshock luminosity to escape as X-rays (moreso
if the geometry is not spherically symmetric).  This would occur after
the plateau.

The model that matches the observed light curve best in
Figure~\ref{fig:model} is qualitatively very similar to the model for
SN~1998S presented by \citet{chugai01}, with a higher density inner
wind transitioning to a lower density outer wind (again, this drop may
merely reflect a larger escape fraction of X-rays at lower optical
depth or a different geometric covering factor; our simple model
cannot constrain this further).  The radial density profile we adopt
is shown by the black line in the upper-right inset of
Figure~\ref{fig:model}.  Note, however, that this is the wind density
that occupies only 15\% of the solid angle, following our assumptions
about geometry discussed above.  Therefore, this same inset to
Figure~\ref{fig:model} shows a dashed magenta curve, which is what the
model wind density would be if the same mass were averaged over $4\pi$
steradians; this is more appropriate when comparing to the spherically
symmetric wind density in the model for SN~1998S by \citet{chugai01},
shown by the green dashed curve in the same inset.  So, comparing
these one can see that the CSM around PTF11iqb is qualitatively very
similar to that of SN~1998S, but with less mass and with a division
between the inner and outer wind at $\sim10^{15.5}$ cm instead of at
$\sim10^{15}$ cm.

Our adopted CSM has an effective (i.e., spherically averaged) wind
density parameter ($w \equiv \dot{M}/v_w$) of $w=10^{15}$ g cm$^{-1}$
for the inner CSM and $2.5\times10^{14}$ g cm$^{-1}$ for the outer
CSM.  With $\dot{M} = w v_w$, these densities translate to pre-SN
mass-loss rates of roughly $1.5 \times 10^{-4}$\,M$_{\odot}$ yr$^{-1}$
(inner) and $4 \times 10^{-5}$\,M$_{\odot}$ yr$^{-1}$ (outer) for a
wind speed of 100\,km s$^{-1}$.  The total mass in the inner shell is
about $10^{-3}$\,M$_{\odot}$ lost in the $\sim8$ yr before core
collapse, and about 0.04\,M$_{\odot}$ lost in the $\sim1000$ yr before
core collapse (out to $\sim3\times10^{17}$ cm). These are only
order-of-magnitude estimates, owing to possible variation in wind
speed, X-ray escape fraction, and geometry.  The CSM mass is small
compared to the several M$_{\odot}$ in the massive expanding SN
ejecta, but the CSM can be dense enough to resist the expansion of the
SN ejecta if it is in a thin disc that intercepts a small fraction of
the solid angle of the explosion.  These pre-SN mass-loss parameters
are less extreme than for other SNe~IIn
\citep{kiewe12,svirski12,ofek13b}, and are within the range of
mass-loss rates of several observed classes of moderately massive
evolved supergiants (RSGs with dense winds like OH/IR stars, YSGs,
B[e] supergiants, and relatively low-luminosity LBVs; see
\citet{smith14} for a general review of mass-loss rates in evolved
massive stars).  The CSM for PTF11iqb does not require the most
extreme levels of LBV-like eruptive mass loss, but it does imply some
significant episodic ejection or wind modulation in the decade before
core collapse.  Both PTF11iqb and SN~1998S had extended CSM consistent
with RSG winds, although on the high end of known mass-loss rates for
RSGs \citep{ms12}.

In this model, the total energy radiated by CSM interaction in the
first 100 days is at least 10$^{49}$ ergs, although the early rise may
be caused partly to shock-breakout luminosity reprocessed by the dense
inner wind (see, e.g., \citealt{ofek10}).  This is the extra
luminosity that is needed in addition to a normal SN~II-P light curve.
This extra energy can be reduced somewhat if the underlying plateau SN
is more luminous (although it can be at most a factor of $\sim1.6$
more luminous to not exceed the observed day 80--100 luminosity).  The
amount of extra radiated energy in only the first 20 days indicated by
optical photometry is about $7\times10^{48}$ erg, but this ignores a
bolometric correction and so the true radiated energy in this time is
probably a factor of several higher, given the high $\gtrsim$25,000 K
temperature at the earliest times.  In any case, this energy is
several times higher than the typical energy of a UV flash associated
with shock breakout in a RSG without a dense inner wind.

\begin{figure}
\includegraphics[width=3.1in]{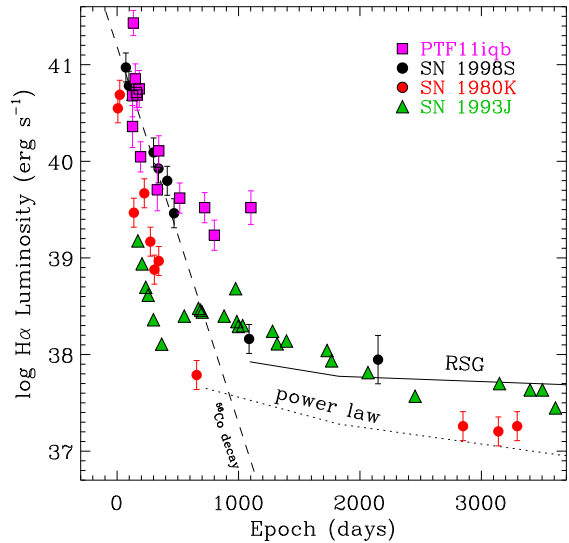}
\caption{Late-time H$\alpha$ luminosity of PTF11iqb, as compared to
  other SNe with strong late-time H$\alpha$ from CSM interaction. We
  include SN~1998S and SN~1980K from Figure 4 of \citet{ms12}, with
  SN~1980K H$\alpha$ data from \citet{dan12}.  H$\alpha$ luminosities
  for SN~1993J are compiled by \citet{chandra09}.}
\label{fig:l8ha}
\end{figure}

\begin{figure*}
\includegraphics[width=4.6in]{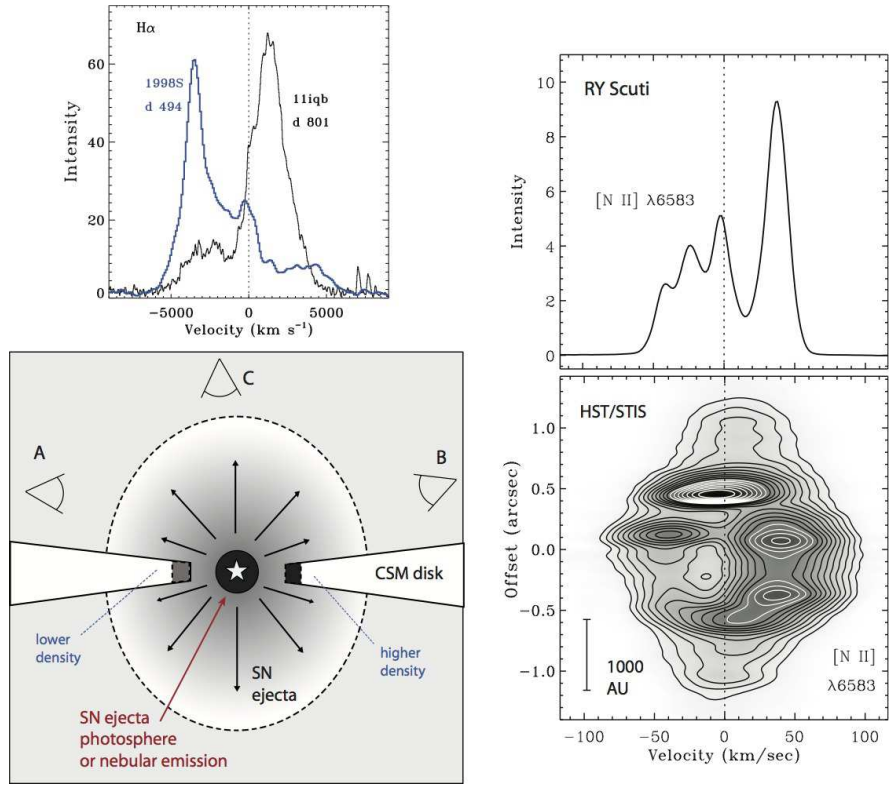}
\caption{These panels relate to the discussion of asymmetry in \S 6 of
  the text.  The upper-left plot compares the late-time asymmetric
  H$\alpha$ profiles of SN~1998S and PTF11iqb.  The lower-left panel
  is a sketch of CSM interaction for a SN running into a flattened
  disc-like CSM that has a lower density on the left side and a higher
  density on the right.  A high-inclination observer at position A
  would see stronger redshifted peaks, as in PTF11iqb, whereas an
  observer seeing the same event from position B would see a
  blueshifted peak as in SN~1998S.  An observer at low inclination
  (position C) would see a narrower and more symmetric line profile.
  The right panels correspond to observations of the torus around the
  mass-transferring eclipsing binary RY~Scuti, from \citet{smith02},
  as seen in [N~{\sc ii}] $\lambda$6583 emission.  The botton panel
  shows the spatially resolved position-velocity diagram ({\it
    HST}/STIS spectra) with the slit aperture running through the
  major axis of the torus.  This reveals an azimuthally asymmetric
  density distribution around the torus, with higher densities on the
  receding side.  The top-right panel shows the integrated line
  profile from this circumstellar nebula, seen in a ground-based
  echelle spectrum.  The radius of the torus is about 1000 AU, similar
  to the CSM encountered by PTF11iqb at late times, and the deduced
  CSM mass is similar as well.  It seems likely that if a SN were to
  expand into a torus like that around RY~Scuti, it would produce CSM
  interaction signatures much like those observed in PTF11iqb at late
  times.}
\label{fig:sketch}
\end{figure*}

\section{Late-time H$\alpha$ and Asymmetry}

After about 100--120 days, corresponding to the drop in continuum
luminosity of the underlying plateau SN light curve, the spectrum of
PTF11iqb took on a different character, dominated by a nebular ejecta
spectrum plus a very strong H$\alpha$ line that we attribute largely
to CSM interaction.  As noted above, the multipeaked H$\alpha$ line
profile is qualitatively very similar to that of SN~1998S
\citep{leonard00,fassia01,pozzo04}, which was interpreted as arising
from CSM interaction in a flattened disc. We note that the late-time
H$\alpha$ from PTF11iqb is also reminiscent of the late H$\alpha$
emission from the nearby SN~IIb explosion SN~1993J
\citep{matheson00a,matheson00b}, where the H$\alpha$ was also
attributed to CSM interaction in an extended disc.

Figure~\ref{fig:l8ha} shows the temporal evolution of the integrated
H$\alpha$ line luminosity in PTF11iqb and in a few other SNe with
strong late-time H$\alpha$.\footnote{Note that at late times, some of
  the spectra are contaminated by very narrow nebular H$\alpha$ and
  [N~{\sc ii}] emission from adjacent H~{\sc ii} regions, where
  variable amounts are present in the spectra owing to seeing
  differences.  We subtracted the contribution of these narrow H~{\sc
    ii} lines from the integrated line flux.} For the first $\sim500$
days, PTF11iqb decays at roughly the $^{56}$Co rate, and is quite
similar to SN~1998S. It is during this time frame that PTF11iqb and
SN~1998S both show a strong blueshifted peak in H$\alpha$, along with
strong nebular lines like the broad Ca~{\sc ii} IR triplet.  During
this time, the H$\alpha$ luminosity may be powered largely by
radioactive decay.  As such, the velocity asymmetry in the H$\alpha$
lines (especially the blue bump), could indicate a highly nonspherical
distribution of $^{56}$Ni.

After about 500 days, however, the H$\alpha$ from PTF11iqb decays more
slowly than in SN~1998S, SN~1993J, and SN~1980K, and no longer traces
the $^{56}$Co radioactive-decay rate.  During these later times,
H$\alpha$ is therefore most likely powered by CSM interaction.  It is
during these same later epochs after day 500 when PTF11iqb's H$\alpha$
profile switches from a blueshifted peak to a stronger redshifted
peak.  It will be interesting to see how PTF11iqb continues to evolve.
In SN~1998S, the drop in H$\alpha$ luminosity reached a floor at
around day 1000, after which it followed a trajectory consistent with
CSM interaction in a very dense RSG wind \citep{ms12}, and SN~1980K
and SN~1993J behaved similarly \citep{fesen99,dan12,matheson00b}.  The
stronger day 500--1100 H$\alpha$ luminosity in PTF11iqb could indicate
an earlier onset of this behaviour in a wind that is an order of
magnitude more dense (which would be extreme, given that the wind
parameters for SN~1998S resembled those of VY~CMa, the strongest known
RSG wind), or perhaps it is merely overtaking a shell or terminal
shock in the RSG wind that temporarily enhances the strength of
H$\alpha$.  Continued observations of PTF11iqb are encouraged.  In any
case, the fact that PTF11iqb had weaker CSM interaction at early times
and stronger interaction at late times compared to SN~1998S suggests
that the pre-SN mass loss in SNe~IIn does not necessarily follow a
common recipe, and that detailed observations of the shock progressing
through the distant CSM may help unravel their highly varied mass-loss
histories.  Most SNe are not observed sufficiently late to see this
behaviour, and/or they are too distant.  If pre-SN mass loss is
punctuated by stochastic mass ejections over a range of timescales,
this would be quite relevant for diagnosing the driving mechanism in
the final phases before core collapse.


Similarly, if the stochastic mass loss is highly asymmetric
(especially azimuthally asymmetric), this may also be an important
clue to its underlying physical mechanism.  When the asymmetric
blueshifted H$\alpha$ line profiles were seen at late times in
SN~1998S, it was inferred that dust obscuration may have played an
important role in making an otherwise more symmetric double-peaked
line appear heavily blueshifted \citep{leonard00,gerardy00,pozzo04}.
\citet{pozzo04} mentioned the possibility that dust formed in the
postshock region (the CDS) may have been contributing to the
asymmetry, as was seen later in the unambiguous case of SN~2006jc
\citep{smith08b}.  However, as noted earlier, we cannot rely upon dust
or ejecta obscuration to account for the strong asymmetry seen in
PTF11iqb, because this time it is the {\it blueshifted} material on
the near side that is missing.  The fact that emission from the far
side of the CSM interaction is brighter provides a strict requirement
that the density distribution in the CSM is intrinsically asymmetric.
Interestingly, the velocity of the brighter red peak is slower
(roughly $+$1100 km s$^{-1}$) as compared to that of the blue peak
(roughly $-$2000 to $-$3000 km s$^{-1}$) on day 801, so stronger
deceleration upon running into denser CSM is an intuitively plausible
explanation for the brightening of the red side of the line.  If
SNe~IIn typically have such asymmetry, then a range of viewing angles
could easily explain why some have redshifted peaks, some have
blueshifted peaks, and others have narrower and centrally symmetric
H$\alpha$ profiles at late times, as depicted in the left panels of
Figure~\ref{fig:sketch}.

What sort of process could produce a flattened CSM geometry that is
significantly more dense on one side than the other?  Axisymmetric
structures in the CSM (discs and bipolar nebulae) may conceivably
arise from rapid rotation in single stars, but high degrees of {\it
  azimuthal} asymmetry are difficult for single stars to achieve.  A
way to have one-sided CSM while still satisfying the equatorial mass
distribution suggested by spectropolarimetry of SN~1998S is to have
mass loss in a binary with some nonzero eccentricity, or with unsteady
mass loss.  As an illustrative example, the right panels of
Figure~\ref{fig:sketch} show observed emission from the nebula around
the Galactic massive star RY~Scuti.  RY~Scuti is a rare example of a
massive eclipsing binary system caught in the brief phase of mass
transfer, where one star is being stripped of its H envelope on its
way to the WR phase, and it is so far the only such system known with
a spatially resolved toroidal CSM nebula
\citep{grundstrom07,smith02,smith11b}.  Figure~\ref{fig:sketch} shows
the spatially resolved velocity structure of the ring and the
integrated emission-line profile of [N~{\sc ii}] $\lambda$6583, from
\citet{smith02}.  The integrated line profile from the nebula is
multipeaked and very asymmetric, with a brighter red peak --- very
much like PTF11iqb at late times, but with slower (preshock CSM)
expansion speeds of only $\pm40$ km s$^{-1}$.  When emission from
RY~Scuti's torus is spatially and spectrally resolved (see the
position-velocity diagram in the bottom right in
Figure~\ref{fig:sketch}), it is clear that the density distribution in
the expanding torus is azimuthally asymmetric, with several clumps
around the torus and generally higher density on the far side
\citep{smith02}.  This azimuthal asymmetry arises despite the fact
that the central eclipsing binary system appears to be tidally locked
and has a circularised orbit \citep{grundstrom07}.  In this case, the
azimuthal asymmetry may arise from episodic mass loss.  Proper motions
of the multiple shells around RY~Scuti reveal two separate major
ejections in the last $\sim250$ yr \citep{smith11b}.  Such sequential
mass-loss episodes in a binary might provide the complex density
structure around PTF11iqb; interestingly, the total mass inferred for
RY~Scuti's nebula of 0.001\,M$_{\odot}$ \citep{smith02} within
$\sim1000$ AU is in the right ballpark.  We infer that if a Type II-P
explosion was surrounded by RY~Scuti's nebula, the resulting CSM
interaction could produce the late-time H$\alpha$ emission seen in
PTF11iqb.  The analogy can only be pushed so far, however, because
RY~Scuti is a binary of an O star and a B supergiant, so it is
unlikely to produce a plateau light curve if it exploded tomorrow.
Nevertheless, it provides a nice illustration of the asymmetry that
can arise in the CSM around interacting binaries.  Many other
possibilities exist to create azimuthally asymmetric structure, such
as wind collisions or other interactions in binary systems with
nonzero eccentricity.

\begin{figure}
\includegraphics[width=3.1in]{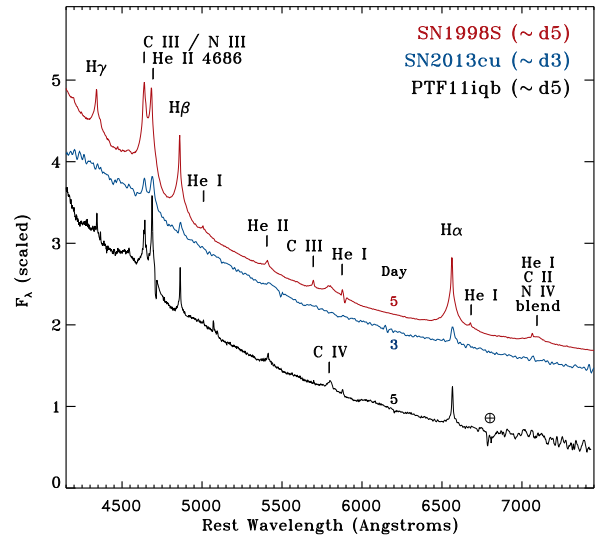}
\caption{Early-time spectra of PTF11iqb (day 2 after discovery,
  roughly day 5 after explosion), SN~2013cu (day 3 after discovery and
  explosion; \citealt{galyam14}), and SN~1998S (day 4 after discovery,
  day 5 after explosion; \citealt{leonard00}) showing the WR features.
  The days noted next to each SN name are approximate times after the
  inferred explosion time.  Various lines of highly ionised species
  such as He~{\sc i}, He~{\sc ii}, N~{\sc iii}, and C~{\sc iii} are
  seen in the spectra, plus broad Lorentzian wings of Balmer lines.
  These WR-like features appear despite the likelihood that both the
  PTF11iqb and SN~1998S progenitors were RSGs, not WR stars, because
  the dense cool wind becomes highly ionised by shock breakout or CSM
  interaction from the fastest ejecta.}
\label{fig:wr}
\end{figure}

\section{Wolf-Rayet Features in the Early Spectrum}

PTF11iqb now joins a few other SNe (Types~IIn and IIb), which have
been caught early and show the WR features associated with He~{\sc ii}
$\lambda$4686 and strong Lorentzian wings of Balmer lines in their
very early-time spectra. The best-studied cases of this so far include
SN~1998S \citep{leonard00,chugai01} and SN~2013cu
\citep{galyam14,groh14}.  Figure~\ref{fig:wr} shows these features in
detail as they appear in the early spectra of PTF11iqb, SN~1998S, and
SN~2013cu.  The spectrum of SN~2013cu from \citet{galyam14} was
obtained from the WISeREP database \citep{yaron12}.

Recently, \citet{galyam14} reported the detection of these same
features in the first few days after explosion in SN~2013cu, which was
a stripped-envelope Type~IIb event.  Those authors interpreted the
spectral features as resembling a WN6h spectral type.  Based on this
similarity, \citet{galyam14} proposed that the progenitor was a
WR-like star.  This is an important claim, since no WR progenitors
have been detected yet for any stripped-envelope SNe, but relatively
cool YSG progenitors have been directly detected in other SNe~IIb, the
three clear cases being SN~1993J, SN~2011dh, and SN~2013df
\citep{ms09,vandyk13,vandyk14}.  (Interestingly, SN~1993J behaves very
similarly to PTF11iqb and SN~1998S at late times.)

However, in this paper we have shown that these same WR features are
seen in PTF11iqb, and they had also been seen previously in
SN~1998S.\footnote{We note one caveat, related to sensitivity or time
  resolution.  An emission feature at $\sim$7100 \AA \ identified by
  \citet{galyam14} as N~{\sc iv} was strong in the spectrum of
  SN~2013cu taken 15.5 hr after explosion, but the feature faded
  beyond detectability by day 3 (see Figure~\ref{fig:wr}).  This
  feature was also seen in SN~1998S, although \citet{leonard00}
  identified it as a C~{\sc ii} blend.  Unfortunately, our earliest
  spectrum of PTF11iqb is too noisy at these wavelengths to see if the
  feature is present at the same level as in SN~1998S at a similar
  time.} This challenges the WR-like progenitor interpretation
forwarded by \citet{galyam14}, since both PTF11iqb (this work) and
SN~1998S \citep{ms12,shivvers14} have CSM consistent with RSG winds.
In the particular case of PTF11iqb, we see a plateau in the light
curve, indicating that the underlying SN event was a Type II-P
explosion --- this requires that the progenitor was an extended cool
supergiant at the moment it exploded, and not a compact WR star.  From
this comparison alone we cannot rule out the possibility that
SN~2013cu had a WR-like progenitor, but clearly a WR progenitor is not
necessary to yield WR features in a SN.  A YSG progenitor, as has been
detected directly in other SNe~IIb, may provide a suitable alternative
for SN~2013cu.  Indeed, comparing the spectrum of SN~2013cu to
radiative-transfer models, \citet{groh14} concludes that the SN~2013cu
progenitor could not have been a WR star, and instead must have been a
yellow hypergiant (with more extreme mass loss than a normal YSG).
The enhancement of N lines in the spectrum that led \citet{galyam14}
to connect SN~2013cu with a WN star (rather than a WC) star is not
exclusive to WR stars, since many cooler evolved massive stars across
the upper HR diagram show N-enriched atmospheres (in any case, the
strong N lines may be more of a temperature effect than abundance;
\citealt{groh14}).  Nevertheless, early time data such as the spectra
obtained for 2013cu and PTF11iqb provide powerful probes of the
immediate pre-SN mass loss \citep{galyam14}.

We conclude that the WR-like spectrum seen at early times has little
to do with the spectral type of the progenitor before explosion.
Instead, it most likely reflects a wind density that is high enough to
be optically thick at radii of $\sim10$ AU after it becomes
ionised.\footnote{The estimate of $\sim10$ AU in the case of PTF11iqb
  comes from the required photospheric radius, given the observed
  luminosity and temperature we quote here.  A radius of 10 AU is also
  consistent with the radius of the CDS on day 2 in our CSM
  interaction model.  \citet{galyam14} quote a similar approximate
  radius of $10^{14}$ cm.} Interestingly, the implied wind density
requires a mass-loss rate (0.001\,M$_{\odot}$ yr$^{-1}$ for SN~2013cu;
\citealt{groh14,galyam14}) that is orders of magnitude higher than
observed winds of WR stars, and well above the line-driving limit (see
\citealt{smith14}); WR stars do not have such high densities at 10 AU
because their winds are fast.  The required density is more easily
achieved in the slow dense winds of cool supergiants (with lower wind
speeds, a lower mass-loss rate is required for the same $w$). Of
course, cool supergiants normally lack the ionising photons needed to
create the characteristic WR spectrum.  A WNH-like spectrum will be
produced, however, when any dense cool wind is bathed in a huge flux
of hard ionising photons that are capable of fully ionising the wind.
The WN features can be strong even when H is abundant in the
atmosphere; as a consequence, there are many WNH stars that are not at
the same evolutionary stage as He-burning WR stars (see
\citealt{smithconti08,crowther07}).  The requisite blast of hard
photons might come from either the UV flash associated with shock
breakout \citep{galyam14}, or from CSM interaction
\citep{leonard00,chugai01} generated when the fastest outer SN ejecta
begin to crash into the slow inner CSM.  Since this occurs inside the
continuum photosphere, it is difficult to know which process
dominates, but nevertheless, either process requires a similarly
optically thick inner wind.

Lastly, we note that independently, Shivvers et al.\ (2014) draw a
similar conclusion regarding the interpretation of the WR features by
\citet{galyam14}.  Shivvers et al.\ present a high-resolution echelle
spectrum of SN~1998S obtained at very early times when the WR emission
features were seen in that object.  They find a slow wind consistent
with a RSG progenitor, and therefore favour a similar interpretation
that a WR-like spectrum at early times does not necessarily indicate a
WR progenitor.

\begin{figure}
\includegraphics[width=2.9in]{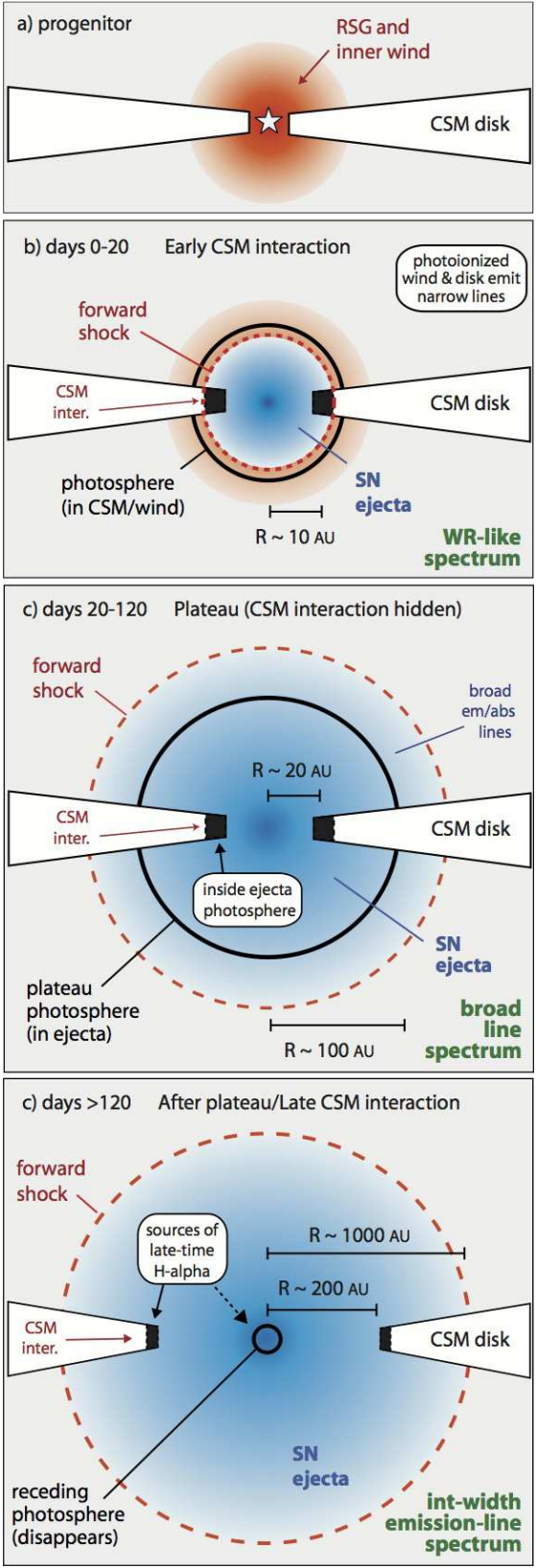}
\caption{Illustration of a proposed sequence of events to account for
  PTF11iqb's evolution.  The four panels ($a$--$d$) are discussed in the
  text, \S 8.}
\label{fig:sequence}
\end{figure}

\section{Overview of PTF11\lowercase{iqb}: Bridging Type
  II{\lowercase{n}} events with normal SN{\lowercase{e}}}

In principle, any type of SN explosion could appear as a Type~IIn,
since the IIn spectroscopic signature is not the result of the
explosion physics, but is caused instead by H-rich (or H-poor/He-rich
in the case of SNe~Ibn) CSM interaction that happens afterward.  When
CSM interaction is strong, the CSM optical depth is high.
Consequently, in cases where CSM interaction substantially enhances
the peak visible continuum luminosity of a SN~IIn, the photosphere can
actually be ahead of the forward shock in the CSM, and hence, it can
effectively mask the identity of the underlying SN ejecta spectrum
(see, e.g., \citealt{smith08} for more details).

Because of its relatively weak CSM interaction at early times,
PTF11iqb affords an interesting opportunity to more clearly see the
underlying SN~II-P photosphere.  The signature of a SN~II-P is seen
from the shoulder in the light curve, which can be explained as a
composite of a normal plateau and CSM interaction, and the
visible-wavelength spectrum toward the end of that plateau matches a
SN II-L, when the broad-lined SN ejecta dominate the observed light.
Moreover, it seems clear from the composite SN~II-P plus CSM
interaction light-curve model that all one needs to do is ``crank up''
the CSM interaction by a factor of a few (in CSM density and
luminosity) to make this object look like the more traditional SN~IIn
explosion SN~1998S.

When we first described the spectral evolution of PTF11iqb in \S 3.2,
we highlighted three main phases:

(1) The early phase (days 0--20) dominated by a smooth blue continuum,
narrow emission lines with broad Lorentzian wings, and the WR-like
spectral features.  During this time, the spectrum was that of an
opaque wind ionised by either CSM interaction or a UV flash from shock
breakout (or both).

(2) The plateau phase (days 20--120) when the spectrum resembled the
broad-lined photospheric spectrum of a normal SN~II-P or II-L, with
broad emission and absorption profiles.  In this phase, the emitted
light is dominated by the fast SN ejecta, as in a normal SN.

(3) Late-time CSM interaction with strong asymmetric H$\alpha$, which
can be subdivided into the early nebular phase (days 120--200) when
H$\alpha$ had a strong blueshifted peak and was probably powered
largely by radioactive decay, and later times (after day 200) when
H$\alpha$ must have been powered by CSM interaction and had a stronger
redshifted peak.

In Figure~\ref{fig:sequence}, we show a sketch that illustrates a
proposed sequence of events to qualitatively explain this observed
evolution of PTF11iqb.  The four panels in this figure are described
below.

{\bf Figure~\ref{fig:sequence}a:} The progenitor is a RSG or yellow
supergiant/hypergiant surrounded by a dense wind and an even denser
disc-like distribution of CSM.  The densest parts of the disc may
reside at a radius of $\sim10$ AU.  The disc is likely the remnant of
some previous binary interaction events.

{\bf Figure~\ref{fig:sequence}b:} For the first $\sim20$ days
immediately after explosion, the SN ejecta crash into the inner dense
wind of the progenitor at a radius of $\sim10$ AU.  However, the wind
is opaque once it is ionised, and a photoionised precursor (due to
either CSM interaction X-rays that are reabsorbed, or to a UV flash
from shock breakout) creates an electron-scattering photosphere
outside the fast SN ejecta.  Narrow lines are emitted by the preshock
wind.  The collision in the equator slows the expansion there, but the
fast ejecta expand freely at other latitudes.

{\bf Figure~\ref{fig:sequence}c:} At days 20--120, the radius of the
SN ejecta photosphere has now grown to $\sim100$ AU at latitudes away
from the equatorial plane, as in a normal SN~II-P.  Although the
photosphere is constantly receding in mass coordinates through the
expanding ejecta, it is larger than the radius where the most intense
CSM interaction is occurring in the equator (10--20 AU). {\it Thus,
  the equatorial CSM interaction region is enveloped by and hidden
  inside the normal broad-lined SN photosphere as seen from most
  viewing angles.}  This is the most likely explanation for the
disappearing or weakening narrow lines in some SNe~IIn.  If CSM
interaction still contributes to the total plateau luminosity, the
photons must diffuse out through the fast ejecta.

{\bf Figure~\ref{fig:sequence}d:} At late times after the plateau drop
(days 100--120), the continuum photosphere recedes to smaller radii
and once again exposes the CSM interaction occurring in the disc.  As
the photosphere recedes completely and the ejecta become optically
thin, the emergent spectrum is dominated by optically thin ejecta
powered by radioactive decay plus CSM interaction in a disc or torus.
The postshock speed of the swept-up CSM is a few 1000 km s$^{-1}$,
indicated by the width of the multipeaked asymmetric H$\alpha$ line.

This sequence of events gives a satisfactory explanation for the
spectroscopic evolution seen in PTF11iqb.  By changing the density and
radial extent of the CSM disc (and perhaps also by changing the
viewing angle of the observer) one can adapt this scenario to other
SNe: with denser or more extended CSM, we get a SN~IIn like SN~1998S,
and with less dense or less extended CSM, we may get normal SNe~II-L
or II-P events.  Thus, this scenario for PTF11iqb provides a direct
bridge between SNe~IIn and SNe~II-P/L.

This, of course, does not mean that all SNe~IIn are actually Type II-P
explosions underneath their CSM interaction, but it does confirm that
{\it some of them} are underlying SNe~II-P that arise from RSGs or
YSGs, although masked at some phases by CSM interaction. Most of the
interest in SN~IIn progenitors has thus far concentrated on LBVs,
partly because in some cases the CSM mass is so high, and also because
the very bright LBV-like progenitors may be easier to detect in
pre-explosion data
\citep{galyam07,gl09,smith14,smith07,smith08,smith10,smith11c}.
\citet{smith09a} suggested that extreme RSGs like VY~CMa have CSM
environments that can produce lower-luminosity SNe~IIn, and
\citet{ms12} showed that the late-time CSM interaction in SN~1998S was
consistent with such a wind operating for $\sim1000$ yr before core
collapse.  More recently, \citet{shivvers14} have also found evidence
for a RSG progenitor for SN~1998S based on very early high-resolution
spectra, as noted above.  There have been other suggestions of RSG
progenitors of some lower-luminosity SNe~IIn as well
\citep{stritz12,smith09b}.

From the point of view of understanding the connection between SNe and
their progenitors, an important question then arises: {\it For how
  long before explosion was PTF11iqb's progenitor a cool supergiant?}
Extreme instabilities in the years leading up to core collapse are
required to eject many M$_{\odot}$ in some cases that produce luminous
SNe~IIn, but there may also be less extreme instabilities that could
drive a wind, cause violent pulsations, or simply inflate the star's
envelope in the final years of a star's life
\citep{sa14,qs12,sq14,ofek14a}.  \citet{yc10} have also suggested that
some luminous RSGs may experience an enhanced superwind phase for
$\sim1000$ yr before core collapse driven by pulsational
instabilities.  During most of its He-burning lifetime, the progenitor
may therefore have looked quite different from how it looked in the
decades or centuries immediately preceding explosion.  Whatever the
physical cause, it does seem clear that PTF11iqb resides between
SN~1998S and normal SNe~II-P in terms of its luminosity and CSM
interaction intensity.

Thus, this type of pre-SN activity, instability, and variation in
stellar structure may play an important role in understanding the
connection between the most common core-collapse events (normal
SNe~II-P) and other objects like SNe~II-L and SNe~IIn that appear to
have some overlap.  It could be that SNe~II-L and SNe~IIn form a
sequence of increasing progenitor mass and increased pre-SN
instability (see, e.g., \citealt{yc10}), or perhaps it could be due to
other factors, such as proximity of a companion star in binary systems
that suffer these instabilities \citep{sa14}.  Exploring the diversity
in CSM density and geometry may be extremely important for sorting
this out, since disc-like CSM around a RSG is unlikely without a
companion star to supply angular momentum.  As noted above, some
objects classified as SNe~II-L, including classic objects such as
SN~1980K, SN~1979C, and others (see \citealt{fesen99,dan12}), do not
have spectra obtained at very early times --- perhaps they would have
been classified as SNe~IIn if they had been caught sufficiently early.
Detecting additional cases like PTF11iqb and understanding how common
they are may help flesh out the continuum of diversity in pre-SN mass
loss.  If there are additional weak SNe~IIn that are not easily
recognised because the SN~IIn signatures are fleeting, it would imply
that pre-SN instability is not limited to only 8--9\% \citep{smith11}
of core-collapse SNe that have been categorised as Type~IIn.

\section*{Acknowledgements}

\scriptsize 

We thank Iair Arcavi, Peter Blanchard, Yi Cao, Ori Fox, Paul Groot,
Asaf Horesh, Michael Kandrashoff, Pat Kelly, Nick Konidaris, Rubina
Kotak, David Levitan, Adam Miller, Yen-Chen Pan, Jarod Parrent, Paul
Smith, and WeiKang Zheng for assistance with some of the observations
and data reduction.  We thank Eran Ofek for helpful discussions and
assistance with the PTF photometric data.
We thank the staffs at Lick, MMT, LBT, Keck, Palomar, and WHT for
their assistance with the observations.  Observations using Steward
Observatory facilities were obtained as part of the observing program
AZTEC: Arizona Transient Exploration and Characterization.  Some
observations reported here were obtained at the MMT Observatory, a
joint facility of the University of Arizona and the Smithsonian
Institution.  This research was based in part on observations made
with the LBT.  The LBT is an international collaboration among
institutions in the United States, Italy and Germany. The LBT
Corporation partners are: the University of Arizona on behalf of the
Arizona university system; the Istituto Nazionale di Astrofisica,
Italy; the LBT Beteiligungsgesellschaft, Germany, representing the
Max-Planck Society, the Astrophysical Institute Potsdam and Heidelberg
University; the Ohio State University and the Research Corporation, on
behalf of the University of Notre Dame, University of Minnesota and
University of Virginia.  The WHT is operated on the island of La Palma
by the Isaac Newton Group in the Spanish Observatorio del Roque de los
Muchachos of the Instituto de Astrofísica de Canarias.  Some of the
data presented herein were obtained at the W. M. Keck Observatory,
which is operated as a scientific partnership among the California
Institute of Technology, the University of California and NASA; the
observatory was made possible by the generous financial support of the
W.M. Keck Foundation. The authors wish to recognise and acknowledge
the very significant cultural role and reverence that the summit of
Mauna Kea has always had within the indigenous Hawaiian community.  We
are most fortunate to have the opportunity to conduct observations
from this mountain.

N.S. received partial support from NSF grants AST-1210599 and
AST-1312221.  E.O.O. is incumbent of the Arye Dissentshik career
development chair and is grateful to support by grants from the
Willner Family Leadership Institute Ilan Gluzman (Secaucus NJ),
Israeli Ministry of Science, Israel Science Foundation, Minerva and
the I-CORE Program of the Planning and Budgeting Committee and The
Israel Science Foundation.  A.G.-Y. is supported by the EU/FP7 via ERC
grant No. 307260, the Quantum Universe I-Core program by the Israeli
Committee for planning and funding, and the ISF, Minerva and ISF
grants, WIS-UK ``making connections,'' and Kimmel and ARCHES awards.
The supernova research of A.V.F.'s group at U.C. Berkeley is supported
by Gary \& Cynthia Bengier, the Richard \& Rhoda Goldman Fund, the
Christopher R. Redlich Fund, the TABASGO Foundation, and NSF grant
AST-1211916.  J.M.S. is supported by an NSF Astronomy and Astrophysics
Postdoctoral Fellowship under award AST-1302771.  K.M. is supported by
a Marie Curie Intra-European Fellowship, within the 7th European
Community Framework Programme (FP7). M.S. acknowledges support from
the Royal Society. D.C.L. and J.C.H. are grateful for support from NSF
grants AST-1009571 and AST-1210311, under which part of this research
(photometry data collected at MLO) was carried out.


\end{document}